\documentclass[onecolumn,11pt]{article}
\usepackage[top=1in, bottom=1in, left=1in, right=1in]{geometry}
\usepackage{amsmath,amsfonts,amscd,amssymb}
\usepackage{graphicx}
\usepackage{epstopdf}
\usepackage{overpic}
\usepackage{cancel}
\usepackage{rotating}
\usepackage{url}
\usepackage{caption}
\usepackage{color}
\usepackage{rotating}
\usepackage{multirow}
\usepackage{wrapfig}
\usepackage{mathtools}
\usepackage{subeqnarray}
\usepackage{setspace}
\usepackage{palatino} 
\usepackage{subcaption}
\usepackage[numbers,sort&compress]{natbib}

\usepackage[bottom,flushmargin,hang,multiple]{footmisc}
\usepackage{lipsum}

\definecolor{header1}{cmyk}{0,0,0,1}

\title{\LARGE{\vspace{-.55in}\textbf{Characterizing Magnetized Plasmas \\with Dynamic Mode Decomposition}}\vspace{-.175in}}

\author{\normalsize{A. A. Kaptanoglu$^{1}$, K. D. Morgan$^2$, C. J. Hansen$^{3,4}$, S. L. Brunton$^5$}\\
\footnotesize{$^1$ Department of Physics, University of Washington, Seattle, WA 98195, United States}\\
\footnotesize{$^2$ CTFusion Inc., Seattle, WA 98195, United States}\\
\footnotesize{$^3$ Department of Aeronautics and Astronautics, University of Washington, Seattle, WA 98195, United States}\\
\footnotesize{$^4$ Department of Applied Physics and Applied Mathematics, Columbia University, New York, NY 10027, United States}\\
\footnotesize{$^5$ Department of Mechanical Engineering, University of Washington, Seattle, WA 98195, United States\vspace{-.2in}}
}
\date{}
\begin{document}
\maketitle
\begin{abstract}
Accurate and efficient plasma models are essential to understand and control experimental devices. 
Existing magnetohydrodynamic or kinetic models are nonlinear, computationally intensive, and can be difficult to interpret, while often only approximating the true dynamics.  
In this work, data-driven techniques recently developed in the field of fluid dynamics are leveraged to develop interpretable reduced-order models of plasmas that strike a balance between accuracy and efficiency.  
In particular, dynamic mode decomposition (DMD) is used to extract spatio-temporal magnetic coherent structures from the experimental and simulation datasets of the HIT-SI experiment.
Three-dimensional magnetic surface probes from the HIT-SI experiment are analyzed, along with companion simulations with synthetic internal magnetic probes.  
A number of leading variants of the DMD algorithm are compared, including the sparsity-promoting and optimized DMD.  
Optimized DMD results in the highest overall prediction accuracy, while sparsity-promoting DMD yields physically interpretable models that avoid overfitting. 
These DMD algorithms uncover several coherent magnetic modes that provide new physical insights into the inner plasma structure. These modes were subsequently used to discover a previously unobserved three-dimensional structure in the simulation, rotating at the second injector harmonic. 
Finally, using data from probes at experimentally accessible locations, DMD identifies a resistive kink mode, a ubiquitous instability seen in magnetized plasmas. \\

\noindent\textbf{Keywords: plasma dynamics, reduced order modeling, dynamic mode decomposition, \\ magnetohydrodynamics, spheromaks} \\

\end{abstract}

\section{Introduction}\label{Sec:introduction}
Understanding and eventually controlling the dynamics of plasmas is essential to the success of fusion reactors, a clean and sustainable energy source based on magnetic confinement.
To reach fusion-level conditions in experiments, a plasma must achieve sustained confinement. However, confinement can be degraded by kink, pressure-interchange, tearing, and ballooning modes as well as a wide array of magnetohydrodynamic (MHD) and kinetic plasma  phenomena~\cite{swanson2003plasma,jetsawtooth,weisen1989mode}. 
Characterizing plasma waves and instabilities is also of considerable interest to the astrophysical plasma physics community, including for understanding electrodynamic coupling between Saturn and its moons~\cite{cassini1,cassini2}, the solar corona~\cite{foullon2011magnetic,antolin2017observational}, and Earth's magnetosphere~\cite{goertz1979magnetosphere}. 
Improved understanding in this field may provide early detection of geomagnetic storms, reducing the economic impact of these events by billions of dollars~\cite{teisberg2000valuation}. 

Despite the importance of modeling and controlling plasmas, the field is challenged by the presence of high-dimensional, nonlinear dynamics that exhibit multi-scale behavior in space and time. 
Despite using a simplified model of the dynamics, plasma simulations often remain very high-dimensional and multi-scale. 
Modeling real-world plasmas often requires computationally-intensive simulations, precluding real-time control of laboratory plasmas. 

This dynamic complexity and high computational cost motivate the development of reduced-order models that map the ambient high-dimensional space to a lower-dimensional \emph{feature} space, where it is possible to model the evolution of dominant spatio-temporal coherent structures.  
Recent studies indicate that over 95\% of the magnetic field energy in an experimental plasma device can be explained by as few as 5-10 spatio-temporal modes, across a large range of parameter regimes, geometry, and degree of nonlinearity~\cite{taylor2018dynamic,pandya,byrne2017study}. 
This implies that the evolution of coherent structures, and possibly the bulk evolution of the plasma, can be understood using a low-dimensional model, with implications for physical discovery, and real-time prediction and control~\cite{Brunton2015amr}. Nearly all fusion-relevant experiments have coherent edge fluctuations and zonal flows with significant effects on transport, including the Alcator C-Mod tokamak~\cite{golfinopoulos2014external}, the DIII-D tokamak~\cite{burrell2002quiescent,DIII-D}, the TJ-II stellarator~\cite{Kobayashi2019}, the High Recycling Steady H-mode in the JFT-2M tokamak~\cite{kamiya2004high}, and the High-Density H-mode in the W7-AS stellarator~\cite{belonohy2008systematic}.

It is thus necessary to identify these low-dimensional patterns from numerical and experimental measurement data. 
The biorthogonal decomposition (BOD) was developed in 1994 for plasmas~\cite{dudok1994biorthogonal}, and has become a standard technique. 
Data-driven modeling is now ubiquitous~\cite{brunton2019data,Brunton2020arfm}, in part because of improvements to the volume, quality, and availability of data. 
The field of reduced-order modeling, i.e. obtaining low-dimensional models to approximate high-dimensional dynamics, has also advanced rapidly in the field of fluid mechanics~\cite{Noack2003jfm,Benner2015siamreview,Kutz2016book,Noack2016jfm,Taira2017aiaa,Carlberg2017jcp,Rowley2017arfm,brunton2019data,Brunton2020arfm}. 

\subsection{Dynamic Mode Decomposition for plasmas}
The dynamic mode decomposition (DMD)~\cite{schmid2010dynamic,Rowley2009jfm,Tu2014jcd,Kutz2016book} is a particularly promising technique, and recently Taylor et al.~\cite{taylor2018dynamic} have successfully applied it to build a sliding-window reduced-order model using only the major magnetic modes in HIT-SI -- the axisymmetric spheromak and a pair of injector-driven modes.  
Their work showed that DMD produces spatio-temporal structures that are more physically interpretable than BOD modes and captures the vast majority of the magnetic energy with a simple rank-3 model in both experimental and simulated HIT-SI discharges, suggesting that the device may be amenable to control~\cite{proctor2016dynamic}. 
Beyond the HIT-SI experiment, DMD has been successfully applied to identify limit-cycle dynamics in 2D turbulent cylindrical plasma simulations~\cite{Sasaki2019}.

DMD is particularly attractive for physics research relevant to plasma waves and instabilities as it decomposes time-series signals into spatially correlated modes that are constrained to have periodic dynamics in time, possibly with a growth or decay rate. Thus, DMD results in a reduced set of spatial modes along with a linear model for how they evolve in time.  
Although DMD yields a linear model, the algorithm has strong connections to nonlinear dynamical systems via Koopman operator theory~\cite{Rowley2009jfm,Mezic2005nd,Mezic2013arfm,noe2013variational,nuske2014jctc,williams2015data,Takeishi2017nips,klus2017data}. 
There have also been several recent innovations and extensions to DMD that improve its ability to model complex systems, including for control~\cite{proctor2016dynamic}, multi-resolution analysis~\cite{kutz2016multiresolution}, nonlinear observations~\cite{williams2015data}, and modal analysis from data that is undersampled in space~\cite{Gueniat2015pof,Brunton2015jcd} and time~\cite{Tu2014ef}. 
Although DMD is known to be sensitive to noise~\cite{Bagheri2014pof}, several algorithms exist to address this issue~\cite{Hemati2014pof,Dawson2016ef,askham2018variable}. 
This collection of DMD algorithms may provide more efficient reduced-order models and deeper physical insight into the inner workings of plasmas.  

\subsection{Contributions of this work}
\begin{figure*}[!tp]
    \centering
    \includegraphics[width=0.98\textwidth]{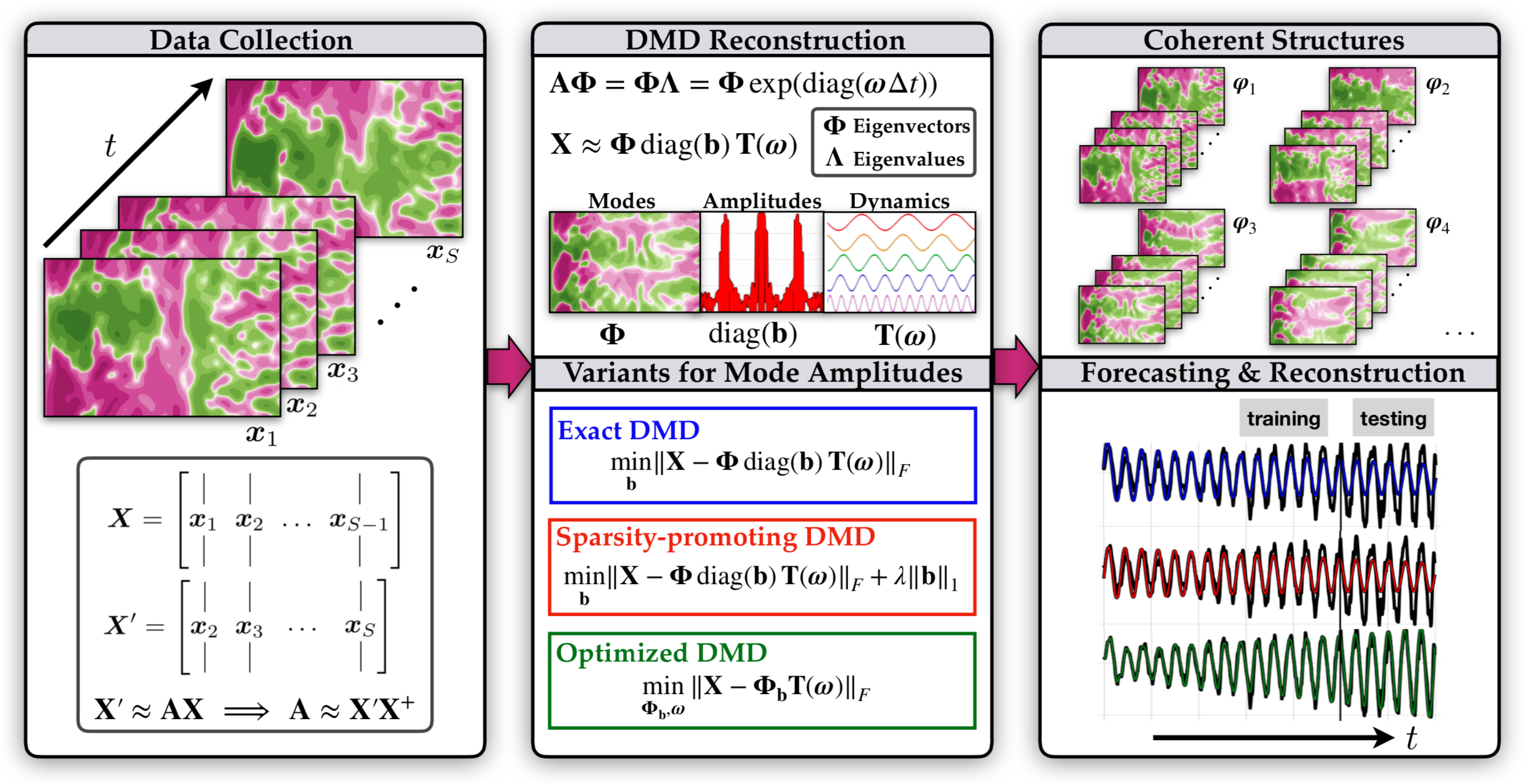}
    \vspace{-.05in}
    \caption{Illustration of the work flow of this paper applied to a set of plasma measurements.}
    \label{fig:workflow}
\end{figure*}

In this paper, we extend the analysis of plasmas via DMD through the use of recent inventions in data-driven discovery~\cite{brunton2019data}. We examine coherent magnetic structures in experimental and simulation HIT-SI data, utilizing all the surface probes and internal magnetic probes. Extensions of the original DMD algorithm, including the sparsity-promoting algorithm of Jovanovic et al.~\cite{jovanovic2014sparsity} and the optimized DMD of Askham and Kutz~\cite{askham2018variable}, lead to improvements in the characterization of physical modes beyond the two major magnetic structures and illustrate the ability of these methods to identify the full spatial and temporal dependence of a common plasma instability. In particular, we identify coherent modes beyond the dominant injector-driven modes and spheromak and, in conjuction with the simulation data, analyze previously unobserved 3D global structures. In addition, we fully characterize a resistive kink instability, illustrating that these methods can be used for instability identification and possibly for real-time control. 
To promote reproducible research and open science, the parallel python code used for this analysis can be found at \url{https://github.com/akaptano/PlasmaPhysics_DMD}.
An overview of the workflow of methods and analysis in this paper is illustrated in Fig.~\ref{fig:workflow}.  

The remainder of the paper is outlined as follows:  Section \ref{Sec:HIT-SI} describes the geometry and experimental configuration of the HIT-SI device. 
In Section \ref{Sec:roms}, the BOD and the three DMD techniques are reviewed. 
Section~\ref{Sec:ExperimentalAnalysis} compares these algorithms on experimental data to illustrate their strengths and weaknesses. 
In Section~\ref{Sec:NumericalAnalysis}, we investigate the magnetic structures observed in a simulation of a large-size version of HIT-SI, named BIG-HIT~\cite{morgan2019formation}. 
The spatial structure of the dominant modes (i.e. the spheromak $f_0$ and the harmonics of the injector frequency $f_1^\text{inj}$, $f_2^\text{inj}$, $f_3^\text{inj}$) is analyzed and connected to physical understanding. 
A resistive (1,1) kink instability is characterized in excellent agreement with the full simulation data and theoretical prediction. 
Finally, comparisons are made between the results obtained with 24 internal magnetic probes, and those made with 5120 internal probes, to illustrate the power of these methods to retain their applicability on a sparse set of measurements. 

\section{The HIT-SI Experiment}\label{Sec:HIT-SI}
The helicity injected torus with steady inductive (HIT-SI) helicity injection experiment~\cite{jarboe2006spheromak} is an experiment investigating current drive and magnetic self-organization for magnetic confinement fusion, with an emphasis on studying formation and steady-state sustainment of a spheromak~\cite{victor2014sustained}. This experiment exhibits coherent magnetic~\cite{hansen2015numerical,taylor2018dynamic} and velocity flow~\cite{hossack2015study,morgan2018finite} structures and therefore is an ideal choice for the application of these methods.

\subsection{Experimental setup}
\begin{figure*}[!tp]
    \centering
    \begin{subfigure}{0.40\textwidth}
    \includegraphics[width=0.9\textwidth]{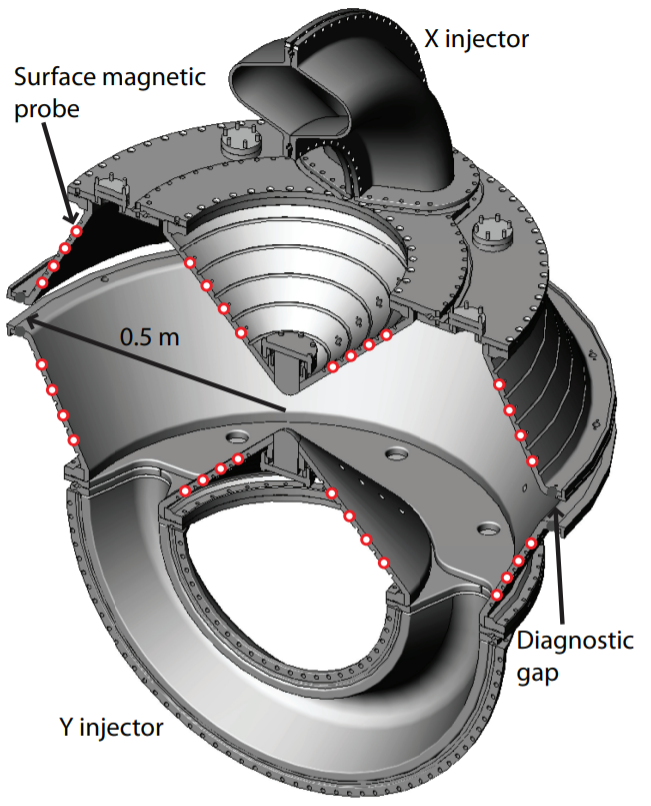}
    \end{subfigure}
    \begin{subfigure}{0.40\textwidth}
    \includegraphics[width=0.95\textwidth]{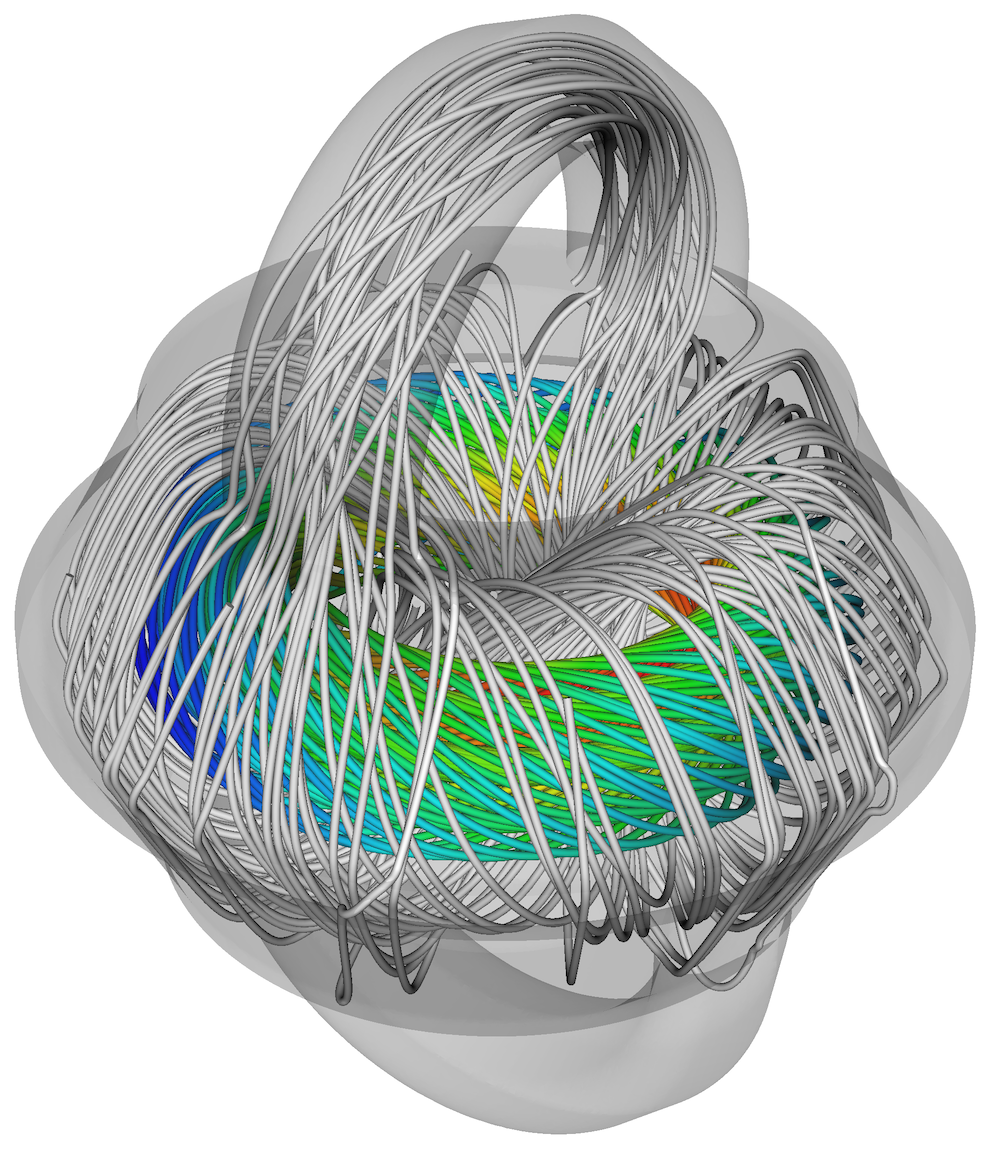}
    \end{subfigure}    
    \caption{Left: A cross section of the device shows the toroidal structure, the two helicity injectors, the surface probes, and the diagnostic gap. Figure reproduced with permission~\cite{victor2015development}. Right: Representative equilibrium during sustainment with an injector shows an axisymmetric spheromak (rainbow) surrounded by field lines tied to the injector (gray).}
    \label{fig:hitsi}
\end{figure*}

\begin{figure}
    \centering
    \includegraphics[width=0.7\textwidth]{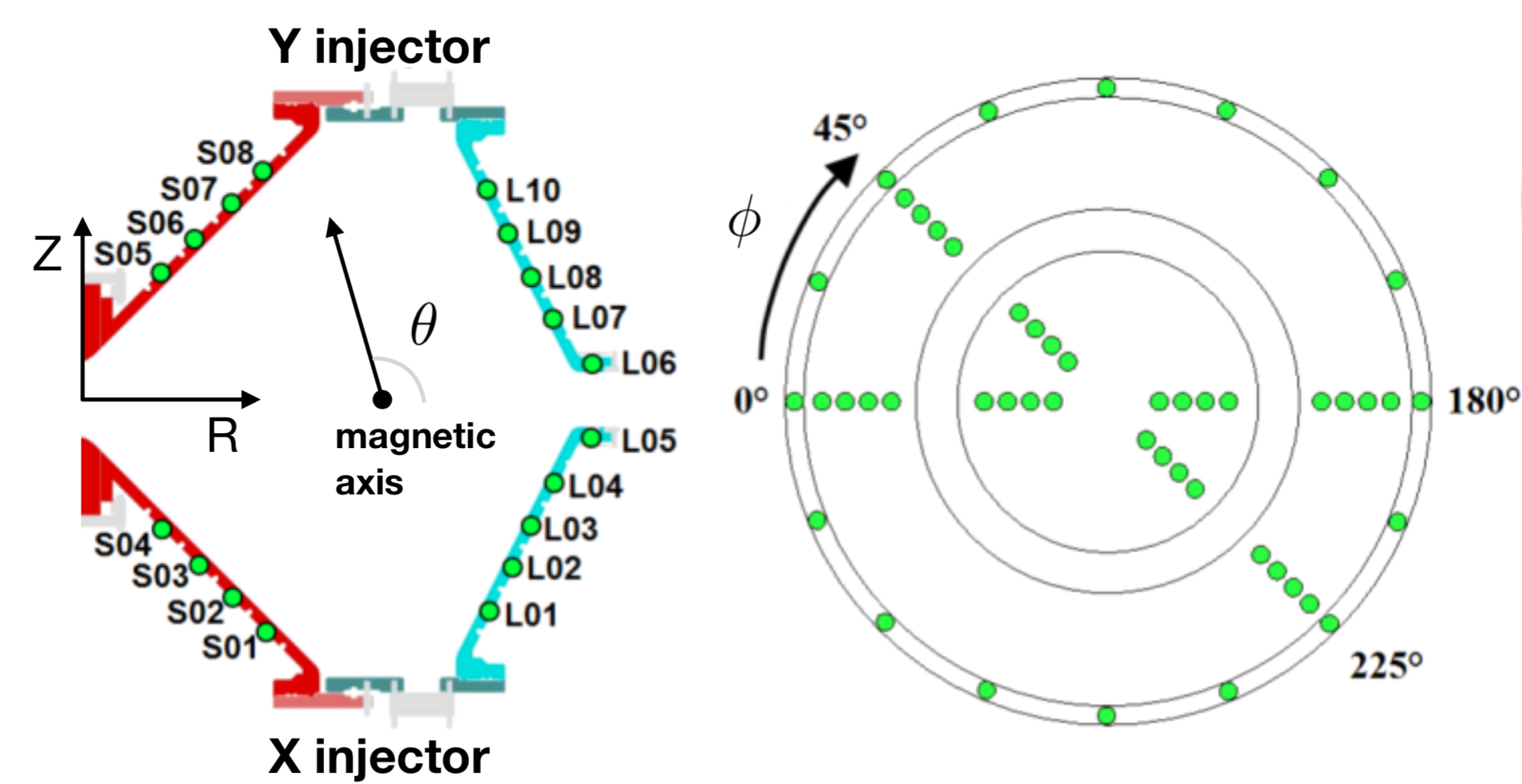}
    \caption{Surface probe locations in a cross section, and from a top view, with 96 probes in total.}
    \label{fig:surfprobe}
\end{figure}

 HIT-SI consists of an axisymmetric main chamber that conserves magnetic flux and two inductive injectors (called the X injector and Y injector) mounted on each end as shown in Fig.~\ref{fig:hitsi}. The HIT-SI experiment has an array of magnetic field probes that encircle four poloidal cross sections at toroidal angles $\phi = $ 0, 45, 180, and 225 degrees, illustrated on the right in Fig.~\ref{fig:surfprobe}. On the left side in Fig.~\ref{fig:surfprobe}, the 18 surface probes are shown in one of the four identical poloidal cross sections. There are also additional probes, labeled $L05$ and $L06$, which are spaced out every $22.5^o$, for a total of 96 probes. Each probe measures the components of the magnetic field, $\boldsymbol{B}$, in the toroidal $\phi$ and poloidal $\theta$ directions, and $B_r \approx 0$ at each probe location. The experimental probes have a time resolution $\Delta t \approx 2$ $\mu s$. 
 
\begin{figure}[!tp]
    \begin{center}
    \begin{overpic}[width=0.6\textwidth]{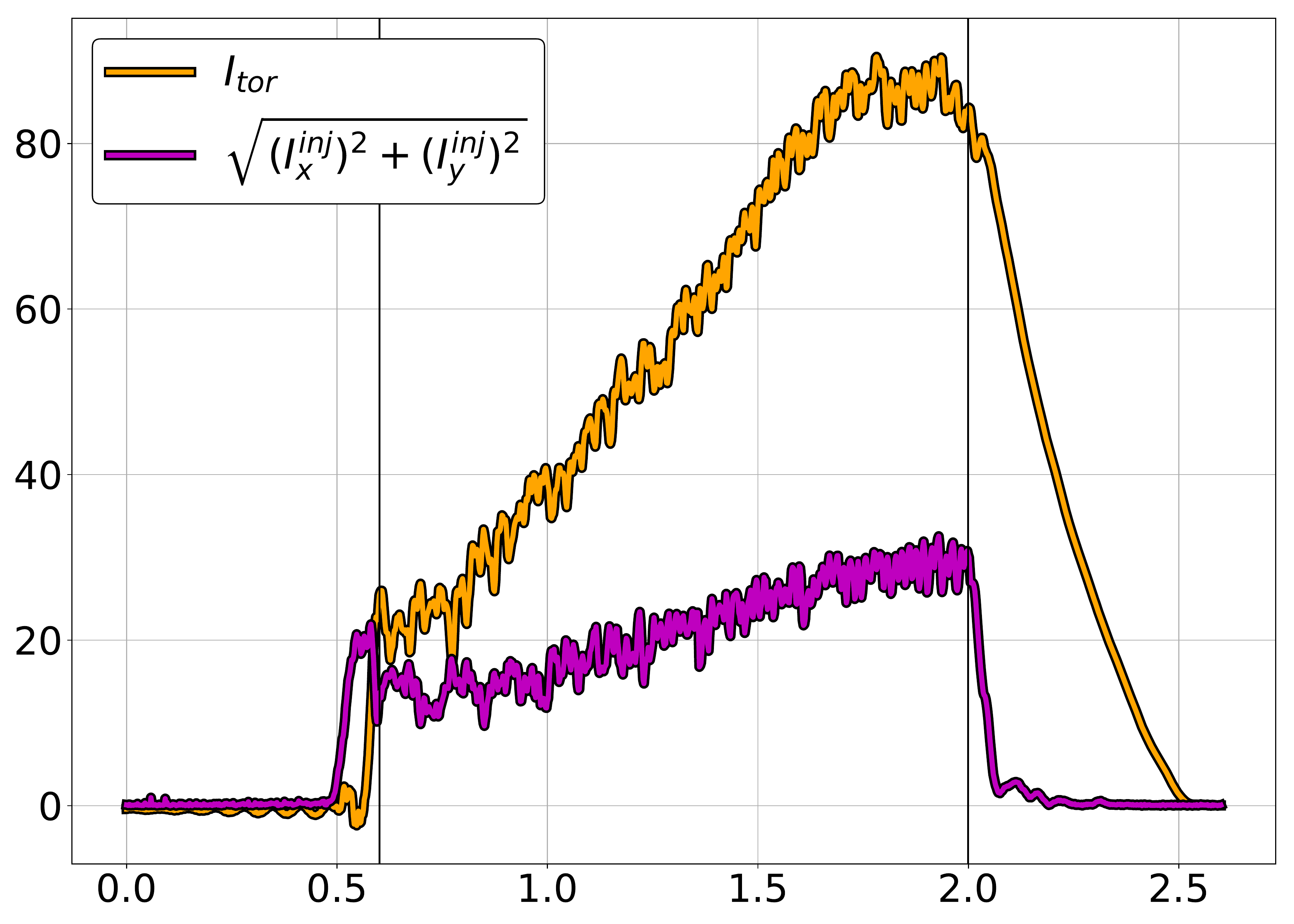}
    \put(45,-4){t (ms)}
    \put(-12,40){$I$ (kA)}
    \end{overpic}
    \end{center}
    \caption{Toroidal current and injector current waveforms for an experimental HIT-SI discharge, with injector frequency $f_1^\text{inj} = 14.5$ kHz. The black vertical lines indicate the sustainment regime when the spheromak has formed and is being sustained by the injectors.}
    \label{fig:itor}
\end{figure}

 Circuits on each injector apply a voltage and an axial flux. These waveforms are oscillated in phase with typical  frequencies $f_1^\text{inj} \approx 5-70$ kHz so that the cumulative power and magnetic helicity injected by the two injectors is approximately constant in time during the discharge. In an experimental shot, the spheromak is created during an initial formation period, and then sustained by the injectors. The analysis presented here will focus on this period of sustainment for all discharges considered in this paper. Because DMD associates the spheromak and the injectors each with a single oscillation frequency, the modes are denoted $f_0$ ($f_0 \approx 0$) and $f_1^\text{inj}$ throughout the paper. The higher harmonics of the injector frequency are  $f_2^\text{inj}$, $f_3^\text{inj}$, and so on. A description of the equilibrium profile and postulated current drive mechanism during sustainment can be found in previous work~\cite{jarboe2012imposed}. The sustainment period of each experimental discharge, indicated by the vertical black lines, is illustrated in Fig.~\ref{fig:itor}. 
 
 In HIT-SI, some of the characteristic time scales overlap, such as the toroidal alfven time and the injector frequency forcing. The magnetic topology is fundamentally 3D and magnetic perturbation amplitudes are comparably large ($|\delta \boldsymbol{B}|$/${|\boldsymbol{B}|} \approx 10 \%$)~\cite{hansen2015simulation}. Because experiments near fusion conditions often have well-separated characteristic scales and much smaller perturbation amplitudes~\cite{wesson2011tokamaks}, improved performance of these methods should be expected on these devices. These machines also typically benefit from considerably more measurement data from a diverse set of high-resolution spatio-temporal diagnostics. 
 
\subsection{Simulations of HIT-SI}
In addition to analyzing experimental data in Section~\ref{Sec:ExperimentalAnalysis}, in Section~\ref{Sec:NumericalAnalysis} we investigate coherent magnetic structures for the BIG-HIT extended MHD simulations using the NIMROD code~\cite{sovinec2004nonlinear}. NIMROD does not simulate the injector geometry and instead imposes boundary conditions on the top and bottom of the main chamber to model the injectors. BIG-HIT is identical to a typical HIT-SI simulation, but the device has been enlarged by a factor of $2.5$. Morgan et al.~\cite{morgan2019formation} provides more details on this simulation. 

Simulations use the experimental surface probes as well as a set of internal probes, all of which measure the magnetic field. In principle any set of measurements normalized to specific units may be used. 
For simplicity, all of the measurements and later results are reported in units of Gauss, and hereafter magnetic field units are omitted. In order to analyze the toroidal structure of the internal magnetic field, 32 internal magnetic probe (IMP) arrays are placed equally spaced toroidally at the axial location $Z=0$, called the midplane. Each array contains 160 measurement points at equally spaced radial locations between $0 \leq R \leq 1.34$ m. It will be shown that a sparse set of only 24 well-separated IMPs captures the mode structures that are observed with the full $160\times 32 = 5120$ IMPs, providing evidence that this analysis is relevant to experimental devices with a small number of unevenly spaced measurements.

\subsection{Data format}
All probe measurements at a fixed time $t_k$ are arranged into a column vector $\boldsymbol{x}_k\in\mathbb{R}^D$, called a snapshot, where $D$ denotes the dimension of the measurement vector, given by the product of the number of spatial probe locations and the number of variables measured at each probe.  
These snapshots are arranged into a matrix $\boldsymbol{X}$
\begin{equation}\label{Eq:DataMatrix}
\boldsymbol{X} = \begin{bmatrix}
    \vline       & \vline & & \vline \\
    \boldsymbol{x}_1 & \boldsymbol{x}_2 & \dots & \boldsymbol{x}_{S} \\ 
    \vline       & \vline & & \vline \\
\end{bmatrix}.
\end{equation}
For both the experimental and simulation data without the internal probes, $D=192$ and $S\approx 500-1000$, as typical discharges are $1-2$ ms with measurement resolution $\Delta t \approx 2$ $\mu$s. The largest simulation dataset has $192$ surface probes and $5120$ internal probes, for a total of $D = 5312$.

\section{Reduced-Order Models}\label{Sec:roms}
Extracting coherent structures from high-dimensional data has been a central challenge in fluid mechanics and plasma physics for decades, but recent advances in data-driven modeling and data volume, quality, and availability, have opened many new possibilities~\cite{Taira2017aiaa,Rowley2017arfm,brunton2019data,Brunton2020arfm}. Reduced-order models exploit these low-dimensional structures by mapping a high-dimensional ambient measurement space to a lower-dimensional \emph{feature} space, where it is possible to describe the evolution of coherent structures. This low-dimensional space is ideal for physical understanding, computational efficiency, and closed-loop control~\cite{Brunton2015amr,Rowley2017arfm}. 
Here we review two leading modal decomposition techniques, the biorthogonal decomposition (BOD)~\cite{dudok1994biorthogonal} and the dynamic mode decomposition (DMD)~\cite{schmid2010dynamic,Rowley2009jfm,Tu2014jcd,jovanovic2014sparsity,Kutz2016book, askham2018variable}. Both methods are broadly applicable to data from simulations or experiments. 

\subsection{Biorthogonal decomposition}
BOD and DMD are both based on the singular value decomposition (SVD), which provides a low-rank approximation of the data matrix $\boldsymbol{X}\in\mathbb{R}^{D\times S}$ from Eq.~\eqref{Eq:DataMatrix}:
\begin{equation}
\boldsymbol{X}= \boldsymbol{U}\boldsymbol{\Sigma}\boldsymbol{V}^*,    
\end{equation}
where $\boldsymbol{U}\in\mathbb{R}^{D\times D}$ and $\boldsymbol{V}\in\mathbb{R}^{S\times S}$ are unitary matrices, and $\boldsymbol{\Sigma} \in \mathbb{R}^{D\times S}$ is a diagonal matrix containing non-negative and decreasing entries.  
The entries of $\boldsymbol{\Sigma}$ indicate how important the corresponding columns of $\boldsymbol{U}$ and $\boldsymbol{V}$ are for describing the structure in $\boldsymbol{X}$.  
In many cases, it is possible to discard small values of $\boldsymbol{\Sigma}$, resulting in a truncated matrix $\boldsymbol{\Sigma}_r\in\mathbb{R}^{r\times r}$, and to approximate the matrix $\boldsymbol{X}$ with only the first $r\ll \min(D,S)$ columns of $\boldsymbol{U}$ and $\boldsymbol{V}$, denoted $\boldsymbol{U}_r$ and $\boldsymbol{V}_r$:
\begin{equation}
\boldsymbol{X}\approx \boldsymbol{U}_r\boldsymbol{\Sigma}_r\boldsymbol{V}_r^*.    
\end{equation}
Throughout the paper $\boldsymbol{V}^*$ is used to denote the complex-conjugate transpose of a matrix $\boldsymbol{V}$, $\bar{\boldsymbol{V}}$ to denote complex conjugation, and $\boldsymbol{V}^+$ to denote the pseudoinverse. 
The truncation rank $r$ is typically chosen to balance accuracy and complexity~\cite{brunton2019data}. 

In practice, the BOD and the SVD are essentially synonymous; in the fluid mechanics community this procedure is called the proper orthogonal decomposition (POD)~\cite{Noack2003jfm,Kutz2016book,brunton2019data}. 
If each column of $\boldsymbol{X}$ corresponds to a set of spatial measurements at a particular time, as above, then the columns of $\boldsymbol{U}$ form an orthogonal spatial basis, referred to as the \emph{topos}, and the columns of $\boldsymbol{V}$ similarly form an orthogonal temporal basis, referred to as the \emph{chronos}.  
 BOD has proven useful for interpreting plasma physics data across a range of parameter regimes~\cite{dudok1994biorthogonal,byrne2017study,galperti2014development}. 
 A typical SVD of HIT-SI surface probes produces three modes that comprise the vast majority of the signal energy, as quantified by the magnitude of the singular values~\cite{hansen2015numerical}. The first mode corresponds to the spheromak equilibrium, $f_0 \approx 0$, while the second and third modes correspond to the injector fields, as they oscillate at approximately the injector frequency $f_1^\text{inj}$. The fourth and fifth modes are sometimes the second harmonics of the injectors, but typically have very low magnitude. 
 Higher modes all appear to correspond to short-lived plasma events and signal noise~\cite{victor2015development}. 
 
Although all of the BOD modes have a dominant frequency, they tend to mix the frequency content, as the SVD/BOD optimization identifies orthogonal modes purely based on energy content, and not to isolate frequency content. This frequency mixing in POD modes was one of the main motivations for the development of DMD in the fluids community.
Other work~\cite{byrne2017study} has shown some promise with the BOD method to investigate kink instabilities, although in a very different experiment and parameter regime than HIT-SI. Previous HIT-SI work using the BOD has shown the identification of plasma-generated instability by subtracting the equilibrium and injector modes from the full reconstruction~\cite{hossack2017plasma}. 

\subsection{Dynamic mode decomposition}
The dynamic mode decomposition is a matrix factorization technique which, like BOD, is based upon the SVD. 
DMD identifies spatial modes that are constrained to have coherent behavior in time, given by oscillations at a fixed frequency, potentially with exponential growth or decay.   
Thus, DMD provides a modal decomposition along with a linear reduced-order model for how these modes evolve in time~\cite{schmid2010dynamic,Rowley2009jfm,Tu2014jcd,Kutz2016book}.  
Identifying modes based on spatio-temporal coherence may avoid the breaking of coherent spatio-temporal structures, especially for plasmas with energetic waves and instabilities which are expected to oscillate, grow, or attenuate with approximately fixed frequency.  
DMD also has strong connections to nonlinear dynamical systems via Koopman operator theory~\cite{Rowley2009jfm,Mezic2005nd,Mezic2013arfm,noe2013variational,nuske2014jctc,williams2015data,Takeishi2017nips,klus2017data}, which is a rapidly advancing field.  
Further, there are several powerful extensions~\cite{proctor2016dynamic,kutz2016multiresolution,williams2015data,Gueniat2015pof,Brunton2015jcd,Tu2014ef,Hemati2014pof,Dawson2016ef,askham2018variable,Towne2018jfm}, making DMD a highly flexible choice that can be tailored for specific scientific goals.  
In this section, we will present the \emph{exact} DMD formulation of Tu et al.~\cite{Tu2014jcd}, which will provide a baseline comparison for the sparsity-promoting~\cite{jovanovic2014sparsity} and optimized~\cite{askham2018variable} DMD algorithms in the following sections.

The nonlinear evolution of the magnetic field may be approximated by a best-fit linear operator $\boldsymbol{A}$ that evolves the state $\boldsymbol{x}_k$ forward in time:
\begin{equation}\label{Eq:DiscreteDynamics}
    \boldsymbol{x}_{k+1}\approx \boldsymbol{A}\boldsymbol{x}_k.
\end{equation}
The dynamic mode decomposition approximates the leading eigenvalues and eigenvectors of the linear operator $\boldsymbol{A}$.
To approximate $\boldsymbol{A}$ from data, we construct two matrices,
$\boldsymbol{X}$ and $\boldsymbol{X}'$ 
\[
\boldsymbol{X} = \begin{bmatrix}
    \vline       & \vline &  & \vline \\
    \boldsymbol{x}_1      & \boldsymbol{x}_2 & \dots & \boldsymbol{x}_{S-1} \\
    \vline       & \vline & & \vline \\
\end{bmatrix}, 
\quad
\boldsymbol{X}' = \begin{bmatrix}
    \vline       & \vline & & \vline \\
    \boldsymbol{x}_2       & \boldsymbol{x}_3 & \dots & \boldsymbol{x}_S \\
    \vline       & \vline & & \vline \\
\end{bmatrix}, 
\]
which are related by
=\begin{equation}\label{eq:LinearDynamics}
\boldsymbol{X}' \approx \boldsymbol{A}\boldsymbol{X}.
\end{equation}
The best-fit linear operator $\boldsymbol{A}$ that satisfies Eq.~\eqref{eq:LinearDynamics} is the solution to the following least-squares optimization:
\begin{equation*}
    \boldsymbol{A} =\underset{\boldsymbol{A}}{\text{argmin}}\|\boldsymbol{X}'-\boldsymbol{A}\boldsymbol{X}\|_F = \boldsymbol{X}'\boldsymbol{X}^+\approx \boldsymbol{X}'\boldsymbol{V}_r\boldsymbol{\Sigma}_r^{-1}\boldsymbol{U}_r^*,
\end{equation*}
where $\boldsymbol{X}^+$ is the pseudoinverse of the matrix $\boldsymbol{X}$. 
However, when the measurement dimension $D$ is large, then $\boldsymbol{A}$ is too large to analyze directly, and instead $\boldsymbol{A}$ is projected onto the first $r$ singular vectors $\boldsymbol{U}_r$:
\begin{equation}
\tilde{\boldsymbol{A}} = \boldsymbol{U}_r^*\boldsymbol{A}\boldsymbol{U}_r =  \boldsymbol{U}_r^*\boldsymbol{X}'\boldsymbol{X}^+\boldsymbol{U}_r = \boldsymbol{U}_r^*\boldsymbol{X}'\boldsymbol{V}_r\boldsymbol{\Sigma}_r^{-1}    
\end{equation}
Next, the eigendecomposition of $\tilde{\boldsymbol{A}}$ is computed:
\begin{equation}
    \tilde{\boldsymbol{A}}\boldsymbol{W} = \boldsymbol{W}\boldsymbol{\Lambda}.
\end{equation}
The diagonal matrix $\boldsymbol{\Lambda}$ contains the eigenvalues $\lambda_j$ of $\tilde{\boldsymbol{A}}$, which are also eigenvalues of $\boldsymbol{A}$.
The corresponding eigenvectors of $\boldsymbol{A}$ may be computed as
\begin{equation}\label{eq:dmd2}
    \boldsymbol{\Phi}= \boldsymbol{X}'\boldsymbol{V}_r\boldsymbol{\Sigma}_r^{-1}\boldsymbol{W}.
\end{equation}
The columns $\boldsymbol{\varphi}_j$ of $\boldsymbol{\Phi}$ are \emph{DMD eigenvectors}  corresponding to \emph{DMD eigenvalues} $\lambda_j$. It is then possible to reconstruct the state at time $k\Delta t$:
\begin{equation}
     \boldsymbol{x}_{k} = \sum_{j=1}^r \boldsymbol{\varphi}_j\lambda_j^{k-1}b_j = \boldsymbol{\Phi}\boldsymbol{\Lambda}^{k-1}\boldsymbol{b}, 
\end{equation}
where $\boldsymbol{b}$ is a vector of DMD mode \emph{amplitudes}.  
In the simplest case, it is possible to approximate $\boldsymbol{b}=\boldsymbol{\Phi}^+\boldsymbol{x}_1$, although the sparsity-promoting and optimized variants below will provide more principled approaches to approximate $\boldsymbol{b}$.  
The data matrix $\boldsymbol{X}$ may then be written as 
\begin{align*}
    \boldsymbol{X} &\approx \begin{bmatrix}
    \vline & & \vline \\
    \boldsymbol{\varphi}_1 & \cdots & \boldsymbol{\varphi}_r \\
    \vline & & \vline
    \end{bmatrix}
    \begin{bmatrix}
    b_1 & & \\
    & \ddots & \\
    & & b_r
    \end{bmatrix}
    \begin{bmatrix}
    1 & \lambda_1& \cdots & \lambda_1^{S-1} \\ 
    \vdots & \vdots & \ddots & \vdots \\
    1 & \lambda_r &\cdots & \lambda_r^{S-1} \\ 
    \end{bmatrix}. 
\end{align*}

The eigenvalues $\lambda_j$ describe the discrete-time dynamical system in Eq.~\eqref{Eq:DiscreteDynamics}.  It is often beneficial to analyze the corresponding continuous-time eigenvalues $\omega_j = \log(\lambda_j)$/$\Delta t$, with $\nu_j = \text{Re}(\omega_j)$/$2\pi$, ${f_j = \text{Im}(\omega_j)}$/$2\pi$. 
It is then possible to approximate the data matrix $\boldsymbol{X}$ as 
\begin{align*}
    \boldsymbol{X} &\approx \underbrace{\vphantom{\begin{bmatrix}
    b_1 & & \\
    & \ddots & \\
    & & b_r
    \end{bmatrix}}\strut \begin{bmatrix}
    \vline & & \vline \\
    \boldsymbol{\varphi}_1 & \cdots & \boldsymbol{\varphi}_r \\
    \vline & & \vline
    \end{bmatrix}}_{\boldsymbol{\Phi}}
    \underbrace{\begin{bmatrix}
    b_1 & & \\
    & \ddots & \\
    & & b_r
    \end{bmatrix}}_{\text{diag}(\boldsymbol{b})}
    \underbrace{\begin{bmatrix}
    e^{\omega_1t_1} & \cdots & e^{\omega_1t_{S-1}} \\ 
    \vdots & \ddots & \vdots \\
    e^{\omega_rt_1} & \cdots & e^{\omega_rt_{S-1}} \\ 
    \end{bmatrix}}_{\boldsymbol{T}(\boldsymbol{\omega})} 
\end{align*}
where $\text{diag}(\boldsymbol{b})$ is a diagonal matrix of the mode amplitudes $b_j$ and $\boldsymbol{T}(\boldsymbol{\omega})$ is a Vandermonde matrix. 
The dynamics of each mode are separated, so that it is possible to isolate and examine a single spatio-temporal structure without the confounding effects of other modes. This will be particularly useful to characterize instability modes. 

It is possible to obtain a better estimate of the mode amplitudes $\boldsymbol{b}$ with the following minimization problem:
\begin{equation}
    \text{argmin}_{\boldsymbol{b}}||\boldsymbol{X} - \boldsymbol{\Phi}\,\text{diag}(\boldsymbol{b})\,\boldsymbol{T}(\boldsymbol{\omega})||_F .
\end{equation}
This formulation is the basis of the DMD extensions presented in the following sections.

\subsection{Sparsity-promoting DMD}
A central tension in reduced-order modeling is that including more modes often increases accuracy while reducing model interpretability. 
However, sparsity promotion through an addition $L_1$ penalty term has become a common technique for machine learning and data analysis~\cite{brunton2019data} because it can produce sparse and interpretable models in terms of a few essential modes. 
Jovanovic et al.~\cite{jovanovic2014sparsity} introduced an $L_1$ penalty in the DMD optimization

\begin{equation}
    \min_{\boldsymbol{b}} \left(||\boldsymbol{X} - \boldsymbol{\Phi}\,\text{diag}(\boldsymbol{b})\,\boldsymbol{T}(\boldsymbol{\omega})||_F + \gamma||\boldsymbol{b}||_{1}\right) 
\end{equation}
to identify the key DMD modes.  They also introduced the more convenient form
\begin{equation}
    \min_{\boldsymbol{b}} \left( J(\boldsymbol{b}) + \gamma ||\boldsymbol{b}||_1 \right),
\end{equation}
where $\gamma$ determines the level of sparsity, and $J(\boldsymbol{b})$ is 
\begin{align*}
     J &= \boldsymbol{b}^*\boldsymbol{P}\boldsymbol{b} - \boldsymbol{h}^*\boldsymbol{b} - \boldsymbol{b}^*\boldsymbol{h} + s, &&   s = \text{Trace}(\boldsymbol{\Sigma}^*\boldsymbol{\Sigma}), \\
         \boldsymbol{P} &= (\boldsymbol{Y}^*\boldsymbol{Y})\circ(\overline{\boldsymbol{T}\boldsymbol{T}^*}),&& \boldsymbol{h} = \overline{\text{diag}(\boldsymbol{T}\boldsymbol{ V\Sigma}^*\boldsymbol{Y})}.
\end{align*}
This nonlinear optimization can be solved with the alternating direction method of multipliers (ADMM)~\cite{ghadimi2014optimal}.

\subsection{Optimized DMD}
Depending on the scientific aims, the absence of a complete set of spatio-temporal DMD modes could be problematic. This lack of completeness implies that reconstructions of specific signals in the data matrix may be less accurate than the BOD. This could be an issue if a very accurate fit of a subset of the data is desired, either for data-driven discovery or for control purposes. 
The optimized DMD of Askham and Kutz~\cite{askham2018variable} addresses this issue by simultaneously considering the best-fit linear operator between all snapshots in time, as opposed to only considering sequential snapshots, as in the standard DMD.  
The optimized DMD results in excellent signal reconstructions, but at the cost of solving a potentially large, nonlinear optimization problem. The key to applying this method is a variable projection algorithm~\cite{askham2018variable} that simplifies the nonlinear optimization. An additional benefit is that snapshots are no longer required to be evenly space in time. Defining $\boldsymbol{\Phi}_b = \boldsymbol{\Phi}\,\text{diag}(\boldsymbol{b})$, the nonlinear minimization problem is now
\begin{equation}
    \min_{\boldsymbol{\omega},\boldsymbol{\Phi_b}} ||\boldsymbol{X} - \boldsymbol{\Phi_b}\boldsymbol{T}(\boldsymbol{\omega})||_F.
\end{equation} 
 
\begin{figure*}[!tp]
\vspace{-.1in}
    \hspace{.25in}
    \begin{subfigure}[t]{0.48\textwidth}
    \begin{center}
    \begin{overpic}[width=\linewidth]{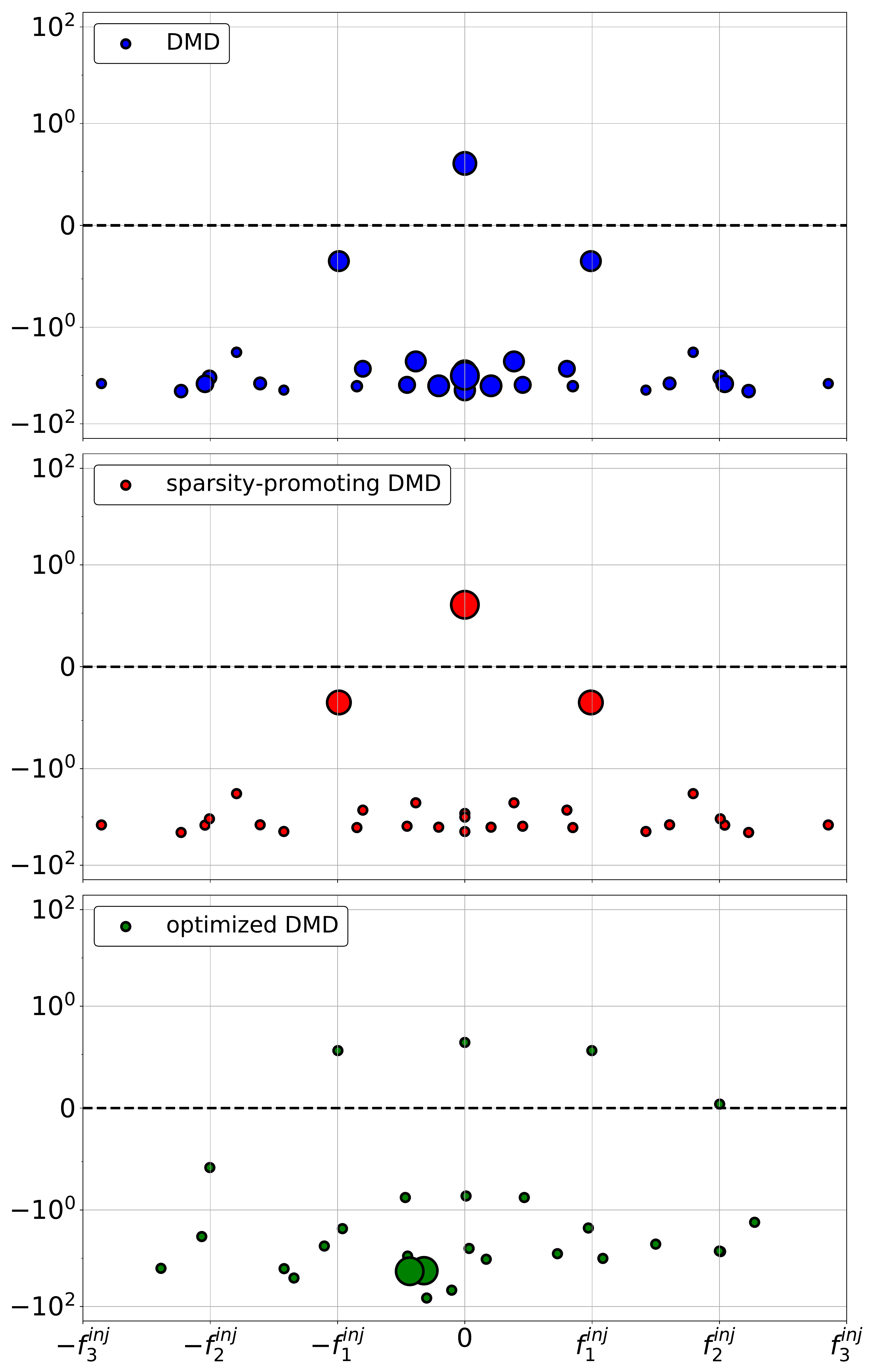}
    \small
    \put(30,-2){$f_j$ (kHz)}
    \put(-9,22){$\nu_j$ (1/ms)}
    \put(-9,54){$\nu_j$ (1/ms)}
    \put(-9,87){$\nu_j$ (1/ms)}
    \end{overpic}
    \end{center}
    \caption{DMD eigenvalues weighted by amplitude.}
    \label{fig:experimentala}
    \end{subfigure}
    \begin{subfigure}[t]{0.48\textwidth}
    \begin{center}
    \begin{overpic}[width=\linewidth]{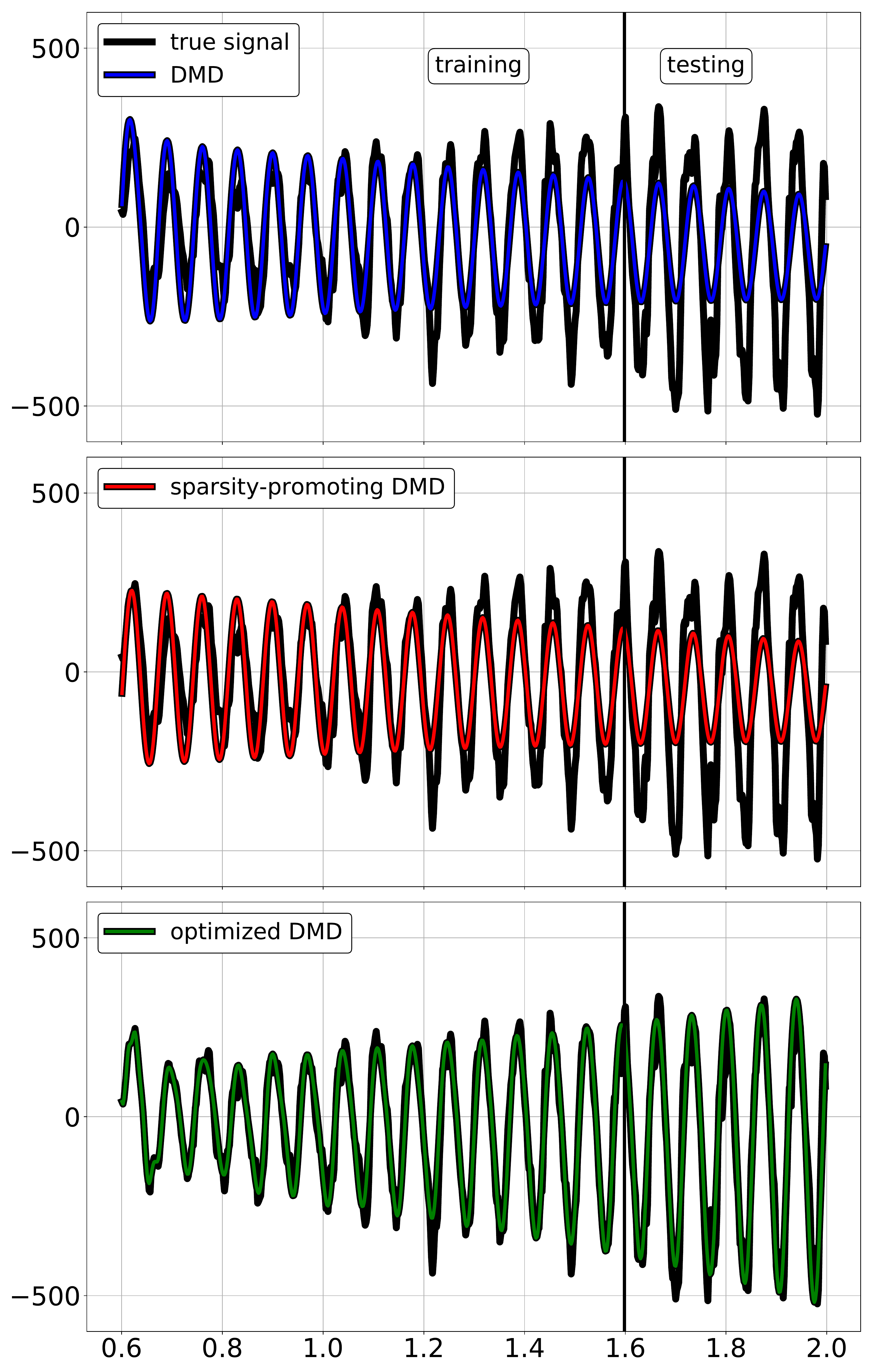}
    \small
    \put(30,-2){t (ms)}
    \put(0,22){$B_\theta$ }
    \put(0,54){$B_\theta$ }
    \put(0,87){$B_\theta$ }
    \end{overpic}
    \end{center}
    \caption{Reconstruction and forecasting of a probe.}
    \label{fig:experimentalb}
    \end{subfigure}
    \caption{(a) The DMD eigenvalues plotted in the complex plane, $\nu_j = \text{Re}(\omega_j)$/$2\pi$ and $f_j = \text{Im}(\omega_j)$/$2\pi$, for the experimental shot at $f_1^\text{inj} = 14.5$ kHz, weighted by $|b_j|$ until some minimum dot size; there are $r=20$ modes. Modes above the dashed horizontal line are unstable. (b) The reconstruction and forecasting performance of each DMD method. The vertical black line indicates where forecasting begins. Optimized DMD provides the most accurate forecast but has growing modes that will eventually diverge.}
    \label{fig:experimental}
\end{figure*}
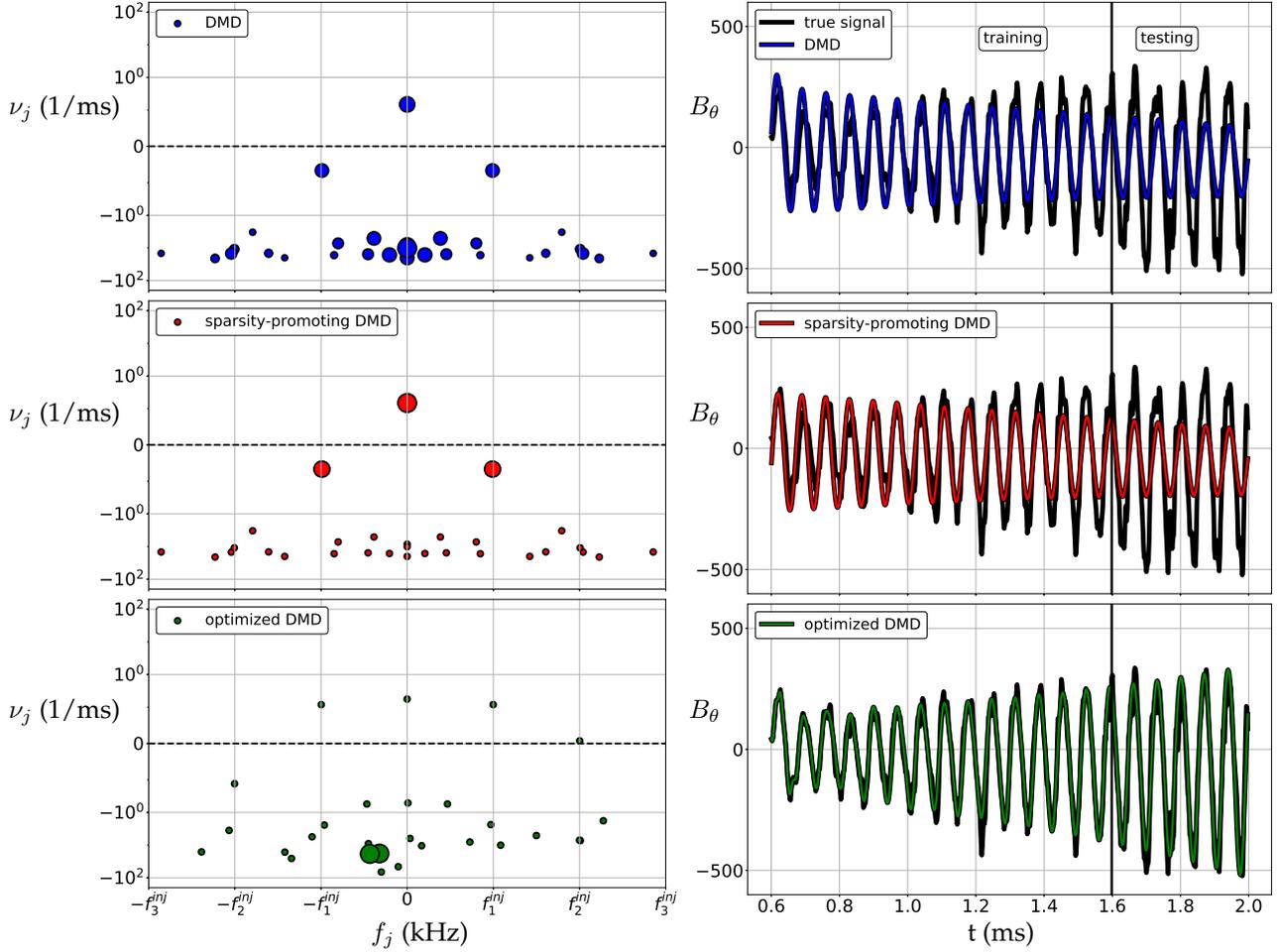

This problem is similar to the original DMD minimization, but the DMD fit is now optimized with respect to both $\boldsymbol{\omega}$ and $\boldsymbol{\Phi}_b$. This problem can be solved with a variable projection followed by the Levenberg-Marquardt algorithm~\cite{levenberg1944method,marquardt1963algorithm}, which relies on a QR decomposition. For speed, the code implements a parallel QR decomposition called direct TSQR~\cite{benson2013direct}.
This algorithm is not guaranteed to find the global minima, but only a local one, so it benefits from an accurate initial guess. 
It is often useful to initialize using the results of the other DMD methods.
Note that for real-valued data, the DMD methods give complex conjugate pairs. The implementation here breaks the complex conjugate symmetry, although it is possible to explicitly retain this symmetry~\cite{askham2018variable}. 
This can be seen visually in Fig.~\ref{fig:experimentala}. Reconstructions are built with the average of each mode and its complex conjugate, guaranteeing real-valued data. 

There is often a trade-off between model interpretability and reconstruction accuracy. 
The sparsity-promoting DMD algorithm produces interpretable models, and the optimized DMD algorithm accurately reconstructs low-energy features and transient instabilities in the data. 
In this way, the strength of using a combination of DMD methods to understand a dynamic system is illustrated in the next sections. Throughout the paper, blue, red, and green are used for the exact DMD, sparsity-promoting DMD, and optimized DMD, respectively.

\section{Comparison of DMD algorithms on an experimental HIT-SI Discharge}\label{Sec:ExperimentalAnalysis}
The exact, sparsity-promoting, and optimized DMD variants are compared on real experimental HIT-SI data in order to understand their relative strengths and weaknesses in identifying interpretable and accurate reduced-order models. 
In the following, an analysis of the high-performance experimental HIT-SI discharge 129499 ($f_1^\text{inj} = 14.5$ kHz), is performed. 
This discharge has been investigated extensively in previous studies~\cite{morgan2018finite,victor2014sustained,hossack2017plasma}.

\subsection{DMD eigenvalues}
The DMD eigenvalues determine the time evolution of the corresponding spatially coherent DMD modes.  
Figure \ref{fig:experimentala} compares the eigenvalues for each of the three DMD methods; the eigenvalues are scaled by their amplitude $|b_j|$. 
In each case, the SVD is truncated at $r=20$ modes to avoid overfitting. 
The $x$-axis represents the imaginary component of the eigenvalue, and the $y$-axis represents the real component so that eigenvalues in the upper half plane are unstable and those in the lower half plane are stable.  
The exact and sparsity-promoting DMD variants result in complex conjugate eigenvalues pairs, which manifests as a symmetry about $f_j=0$.  
However, optimized DMD does not necessarily result in these complex conjugate pairs of eigenvalues. 
The magnitude plot indicates that sparsity-promoting DMD is effective at isolating the three dominant modes, whereas exact and optimized DMD both result in spectra with many energetic modes.  
Thus, sparsity-promoting DMD is capable of extracting and isolating the leading large-scale magnetic structures in the experiment, providing enhanced interpretability. 
In contrast, the next section will show that optimized DMD is needed to extract and analyze small-scale transient modes for a more accurate fit.

\subsection{DMD reconstruction and forecasting}
A common scientific aim is an accurate reconstruction of diagnostic signals using a subset of the modes from a reduced-order model. Decomposing the signal dependence into coherent modes in a low-dimensional space is useful both for physical discovery  and real-time control. 

The advantage of the optimized DMD over the other algorithms is apparent from the reconstruction of a surface probe, as in Fig.~\ref{fig:experimentalb}. 
The exact DMD and sparsity-promoting DMD capture the bulk evolution, but the optimized DMD also captures the deviations. 
The DMD methods are trained on a subset of the data and then evolved in time to forecast the remaining data. 
The optimized DMD provides the most accurate forecast. 
However, this model contains exponentially growing modes that will eventually diverge. 
Including more than $r=20$ modes causes optimized DMD to overfit and results in more unstable modes. 
These observations are reaffirmed quantitatively on simulation data in the next section.

\section{DMD analysis on BIG-HIT simulations}\label{Sec:NumericalAnalysis}
Myriad linear and nonlinear coherent phenomena are observed in both space and laboratory plasmas. Physical understanding can be obtained from reduced-order models by extracting coherent structures, analyzing their temporal frequency content, and decomposing the spatial dependence into Fourier modes. 
The spatial Fourier dependence is important in many experimental devices to determine MHD stability. For toroidal devices, such as HIT-SI, the toroidal and poloidal Fourier wavenumbers are denoted $(n,m)$. In HIT-SI, when the safety factor $q$ satisfies $q > 1$, it can be shown to be kink-unstable to the $(n,m) = (1,1),(2,2),(3,3),...$ modes~\cite{jarboe1994review}. Sawtooth oscillations from these resistive kink modes are also common in toroidal devices when $\text{min}(q) < 1 < \text{max}(q)$~\cite{wesson2011tokamaks,jetsawtooth}.

Here, the DMD methods described in Section III are quantitatively compared based upon their ability to characterize the simulated large-size version of HIT-SI, named BIG-HIT. These 3D simulations were performed using a Hall-MHD model, assuming constant and uniform temperature and density. Relevant constants include the plasma temperature $ = 71$ eV,  density $ = 1.5\times 10^{19}$ m$^{-3}$, resistivity $\eta = 8.9\times 10^{-7}$ $\Omega$m, and injector frequency $f_1^\text{inj} = 14.5$ kHz. For more parameter details, see the original analysis and prior implementations of the model~\cite{akcay2013extended,morgan2019formation}. 

For simplicity, and to demonstrate the ability of these methods to work on small subsets of data, only data in the range $22.7$ ms $\leq t \leq 28.5$ ms is used. The performance on a sparse and spatially well-separated dataset of 24 internal probes is compared with a large and uniformly spaced dataset of 5120 internal probes, in order to illustrate that the conclusions of this analysis on high-resolution data also hold in the limit of far fewer probes.

\subsection{DMD reconstruction error}
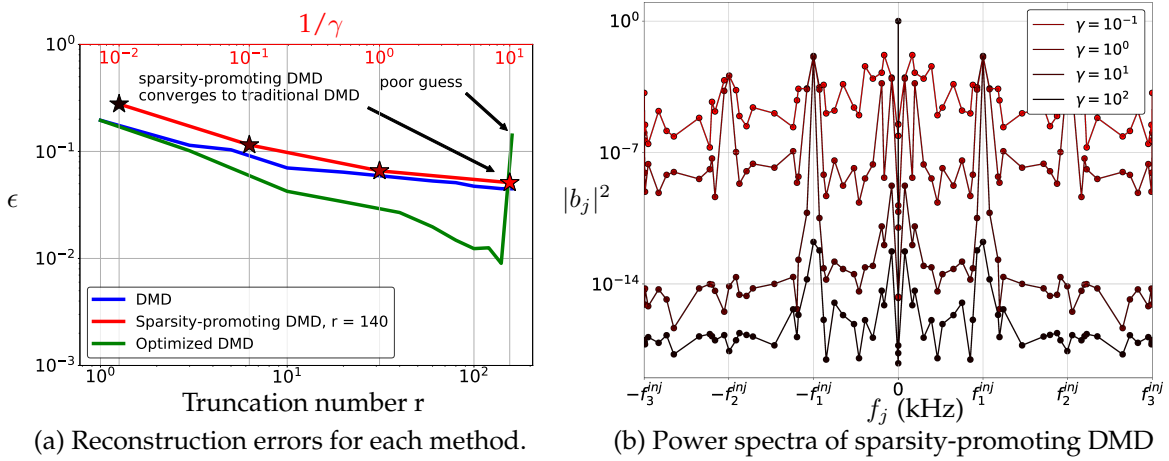
\begin{figure}[!tp]
\centering
     \begin{subfigure}{0.48\textwidth}\hspace{.1in}
    \begin{overpic}[height=.7\textwidth]{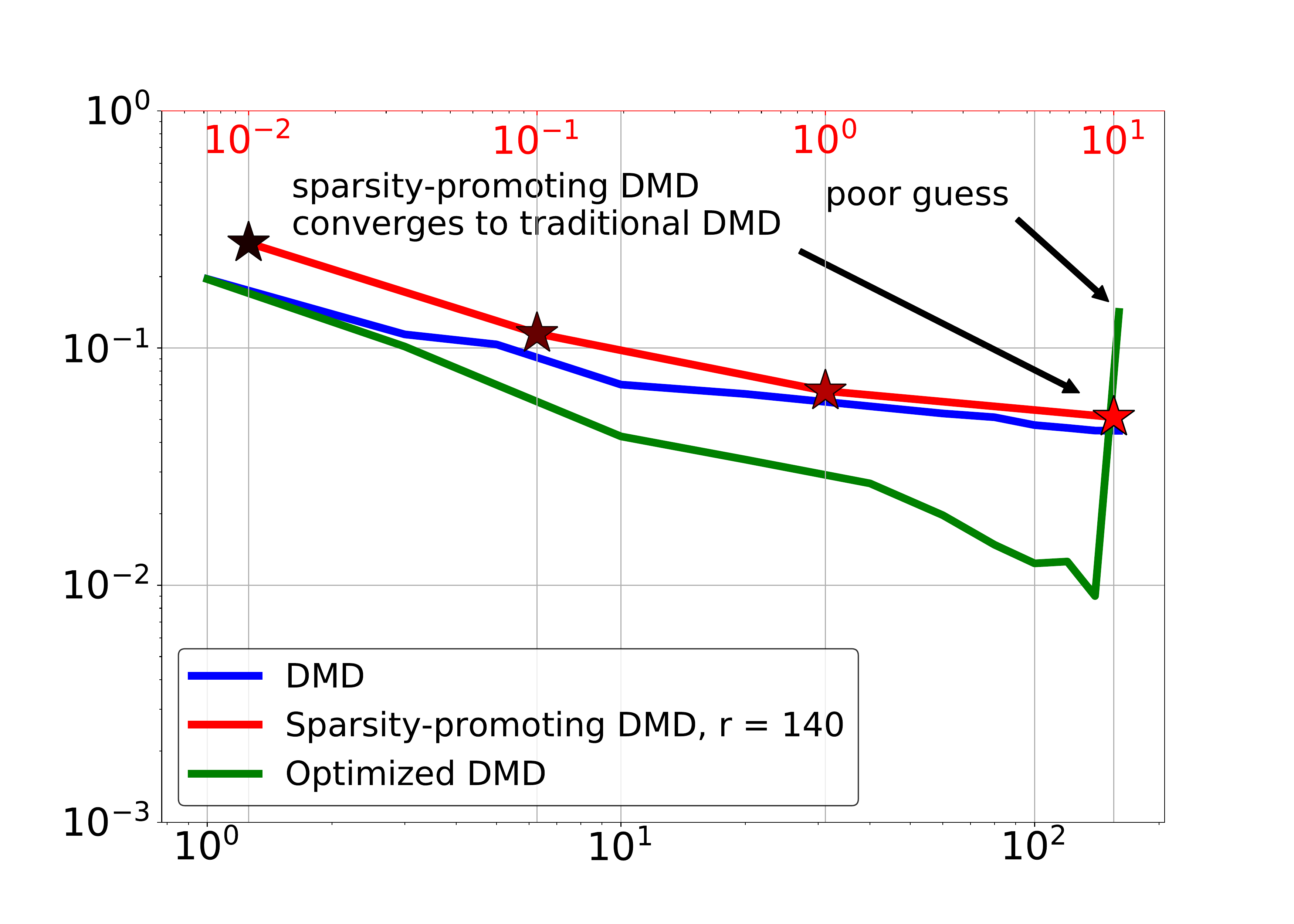}
    \put(0,35){$\epsilon$}
    \small    
    \put(30,0){Truncation number r}
    \put(50,65){\textcolor{red}{1/$\gamma$}}
    \end{overpic}    \caption{Reconstruction errors for each method.}
    \end{subfigure}
    \begin{subfigure}{0.48\textwidth}
    \begin{overpic}[height=.7\textwidth]{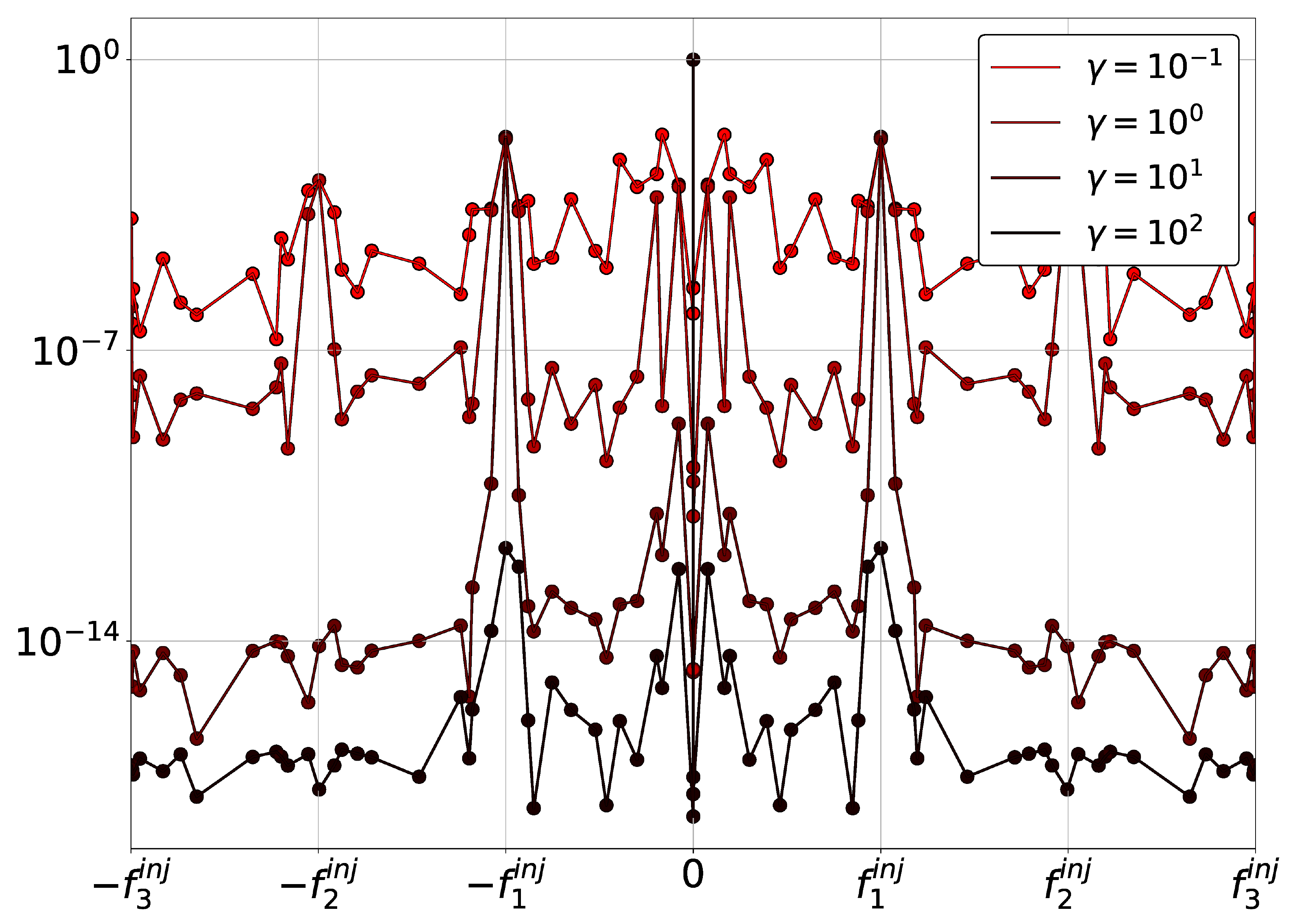}
    \small
    \put(48,-1){$f_j$ (kHz)}
    \put(-4,35){$|b_j|^2$}
    \end{overpic}    
    \caption{Power spectra of sparsity-promoting DMD}
    \end{subfigure}
    \caption{In (a) the optimized DMD obtains the most accurate reconstructions, and sparsity-promoting DMD is shown to converge to exact DMD as $\gamma \to 0$. Stars correspond to the sparsity-promoting DMD power spectra in (b). 
    For large $r$, a bad initial guess results in optimized DMD converging to local minimum with poor reconstruction. In (b) the normalized power spectrum of sparsity-promoting DMD illustrates overall suppression of DMD modes and fewer large peaks as $\gamma$ increases.}
    \label{fig:scaling}
\end{figure}

Each extension of the dynamic mode decomposition has its particular strengths and weaknesses. In the previous section, the sparsity-promoting DMD resulted in interpretable models, while the optimized DMD provided excellent reconstructions of the data.  
Figure~\ref{fig:scaling} provides a quantitative comparison of the different DMD algorithms on the BIG-HIT simulation in the window ${22.7 \text{ ms } \leq t \leq 23.5 \text{ ms}}$, plotting the relative reconstruction error as a function of the SVD truncation rank $r$.  
The relative reconstruction error is defined as
\begin{equation}
    \epsilon = \frac{||\boldsymbol{X}-\boldsymbol{\Phi}\,\text{diag}(\boldsymbol{b})\,\boldsymbol{T}(\boldsymbol{\omega})||_F}{||\boldsymbol{X}||_F}.
\end{equation}
As $\gamma$ increases, the reconstruction error of the sparsity-promoting DMD model also increases.  Moreover, as ${\gamma \to 0}$, this algorithm converges to  exact DMD. 

At $r \approx 140$, the optimized DMD reconstruction error is an order of magnitude smaller than the exact DMD error ($\epsilon \approx 0.009$ against $\epsilon \approx 0.05$). In fact, optimized DMD with $r=10$ obtains the same reconstruction error as exact DMD with $r=140$. However, at $r = 160$, the optimized DMD error increases significantly.  In this case, the initialization procedure chooses a poor first guess consisting of exponentially growing modes, which results in optimized DMD converging to a suboptimal minimum. 
Spurious unstable modes is a general issue with a number of DMD algorithms. These issues are often mitigated with a suitable rank truncation in the SVD (either manually or through sparsity promotion), a different window, or a more accurate initialization procedure.

\subsection{DMD mode characterization}
\begin{figure*}[!tp]
\centering
\begin{overpic}[width=.99\linewidth]{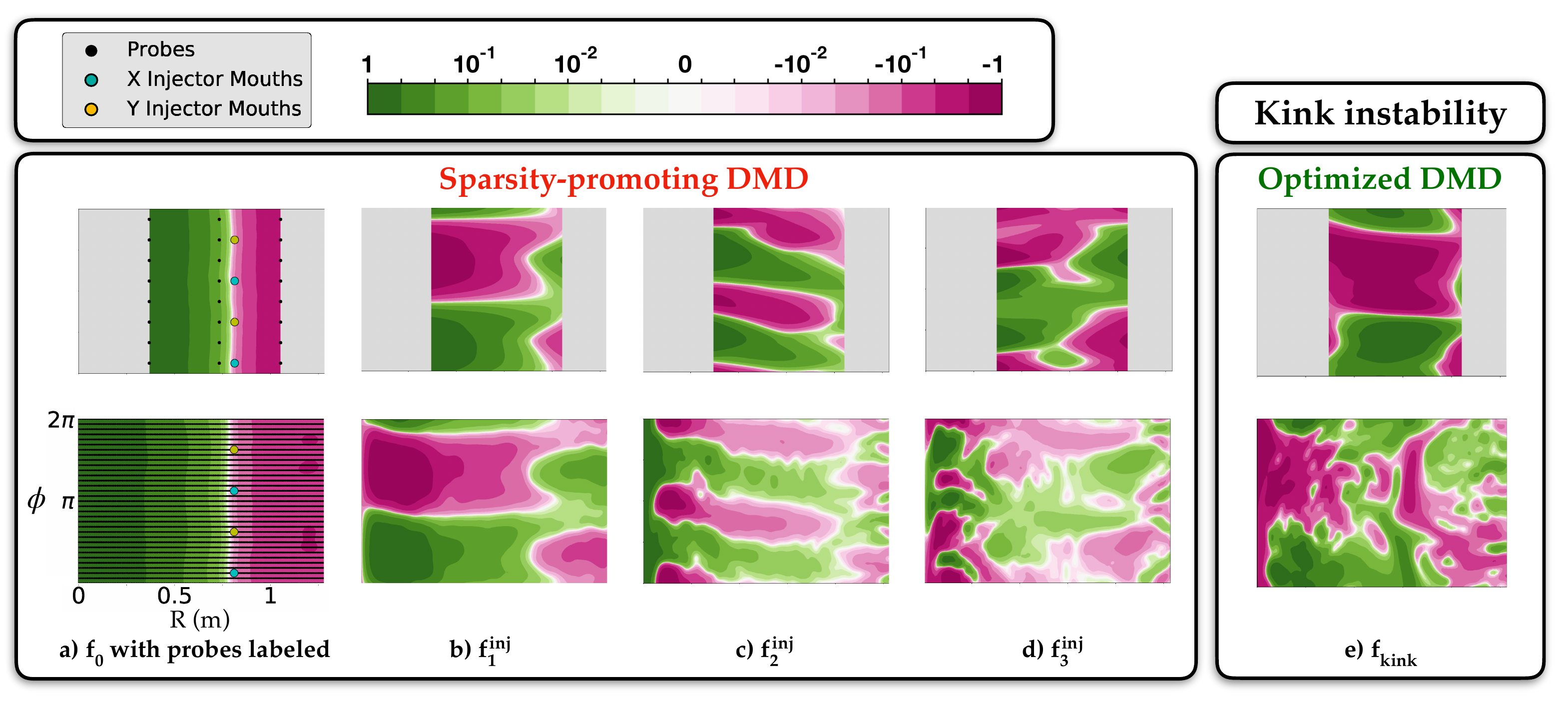}
\end{overpic}
\vspace{-.15in}
    \caption{Normalized $B_z$ at $Z=0$ of the sparsity-promoting DMD modes $f_0,...,f_3^\text{inj}$ and optimized DMD mode $f_\text{kink}$. The small dataset illustrated in the top row has resolution $\Delta R \approx 37$ cm, $\Delta \phi = 45^o$. In the bottom row, $\Delta R \approx 0.8$ cm, $\Delta \phi = 11.25^o$. The sparsity-promoting method captures the vast majority of the spatial structure for each mode even with the small dataset. Fine-scale structure in the kink instability is not captured with the small dataset.}
    \label{fig:contours}
\end{figure*}

Often it is advantageous to obtain a low-dimensional representation of a high-dimensional measurement space in terms of a few modes that capture the dominant coherent structures.  
In this way, physical insights into the dominant structures can be obtained, with the potential for improved understanding or control. In this section,  sparsity-promoting DMD is used to characterize the dominant modes observed in BIG-HIT. 
Here, we will analyze the $B_z$ internal probes at the $Z=0$ midplane, as this is the most analogous quantity to the Poincar\'e plots used to detect kink activity in the simulations in later sections. At $Z=0$, $B_z$ is either parallel or anti-parallel to $B_\theta$. The surface probes, which are not located at $Z=0$, use only $B_\theta$. The sign difference between $B_z$ and $B_\theta$ does not affect the toroidal or poloidal decompositions of the fields; thus, we do not consider this subtlety further for the remainder of the paper. The reconstructions at $Z=0$ for the $f_0$--$f_3^\text{inj}$ and $f_\text{kink}$ modes are illustrated in Fig.~\ref{fig:contours}.

To analyze the spatial structure of each mode, reconstructions of the $B_z$ probe signals are created using only the relevant subset of DMD modes. These reconstructions map the probe locations to the proper location in $(R,\phi)$ space. A Fourier decomposition in the toroidal direction is performed separately for each set of probes with the same radial location. The decomposition
\begin{equation}
    B_z(R,\phi,t) \approx \sum_{n=0}^{N_{max}} 
    \tilde{B}^z_n(R,t)\cos(n\phi-\zeta_{n})
\end{equation}
reveals the toroidal or poloidal structure of the signals, where $2N_{max}+1$ unique toroidal locations are required to resolve the first $N_{max}$ modes. For evenly spaced measurements the coefficients can be found directly by orthogonality. A general method of obtaining the coefficients $\tilde{B}^z_n$ and $\zeta_{n}$ for irregularly spaced angular measurements can be found in appendix F in the reference on HIT-SI surface probes~\cite{wrobel2011study}. The Fourier transform in the poloidal direction is analogous. Taking the absolute value of the  $\tilde{B}^z_n(R,t)$ coefficients and averaging over the Fourier transforms obtained from different radial locations results in
\begin{equation}
    \langle \tilde{B}^z_n(t)\rangle = \frac{1}{N_{rad}}\sum_{i=1}^{N_{rad}}
    |\tilde{B}^z_n(R_i,t)|
\end{equation}
where $N_{rad}$ is the number of radial locations where a separate Fourier decomposition is performed. This quantity gives an average sense of the total toroidal dependence of the reconstructed $B_z$.

Because the surface probes and internal probes have different Z-coordinates, they have separate Fourier decompositions. Only the surface probes admit a poloidal decomposition. For the poloidal Fourier decompositions, the four poloidal arrays of surface probes are separately decomposed and then averaged.

In the previous section, sparsity-promoting DMD was used to capture only the leading order structures, so it is ideal for the analysis of the large coherent magnetic structures oscillating at $f_0$--$f_3^\text{inj}$. This method is particularly relevant for noisy experimental devices to extract coherent modes while avoiding overfitting. 

\subsection{Sparsity-promoting DMD: First injector harmonic}
The HIT-SI injectors drive large magnetic perturbations that sustain the spheromak. They are intentionally operated with an approximate $n=1$ symmetry, but a full picture of the injector field structure is important for understanding the current drive and sustainment in this device. Reconstructions of the magnetic fields with only the injector mode reveal an overwhelming $n=1$ dependence. There is also a phase shift of approximately $180^o$ between the core and edge region of the plasma, shown in Fig.~\ref{fig:contours}, consistent with ion doppler spectroscopy measurements on the experiment~\cite{hossack2015study}. A simulated version of this discharge shows similar results~\cite{morgan2018finite}. 
This suggests a large-scale transition between the inner and outer regions of the plasma, perhaps contributing to a shear layer such as that expected with impose-dynamo current drive~\cite{jarboe2012imposed}. This phase shift occurs at $R\approx 0.8-0.9$ m, close to the closed flux surfaces in the plasma.

While previous work observed this phase shift in the plasma flow velocity, HIT-SI is in a parameter regime where MHD is expected to capture much of the physics. To leading order, in the limit of ideal MHD, electrons and ions are tied to magnetic field lines, and the structure of the velocity and magnetic fields should be similar. However, the motivation for simulating Hall MHD is that this is not quite true; the ion inertial depth is $\approx 6$ cm and the ions are not frozen to the field. If the ion inertial depth effects are small for bulk oscillations, the structures in the velocity and magnetic fields can still be quite similar. 

\subsection{Sparsity-promoting DMD: Second injector harmonics}
Sub-harmonic, harmonic, or nearly-harmonic oscillations are a common feature observed in the nonlinear response to periodic inputs~\cite{khalil2002nonlinear}. Modes oscillating at the harmonics of the injector frequency are often identified by the DMD algorithms. Surprisingly, the DMD mode corresponding to the second harmonic depends mostly on the even toroidal numbers with dominant $n=2$. Moreover, Fig.~\ref{fig:contours} shows that there is a phase shift of approximately $180^o$ at $R \approx 1.05$ m and a smaller shift at $R \approx 0.3-0.4$ m. One possible interpretation is that the toroidally even part of the perturbation is filtered out where there is closed flux.

\begin{figure*}[!t]
\centering
\begin{overpic}[width=.99\linewidth]{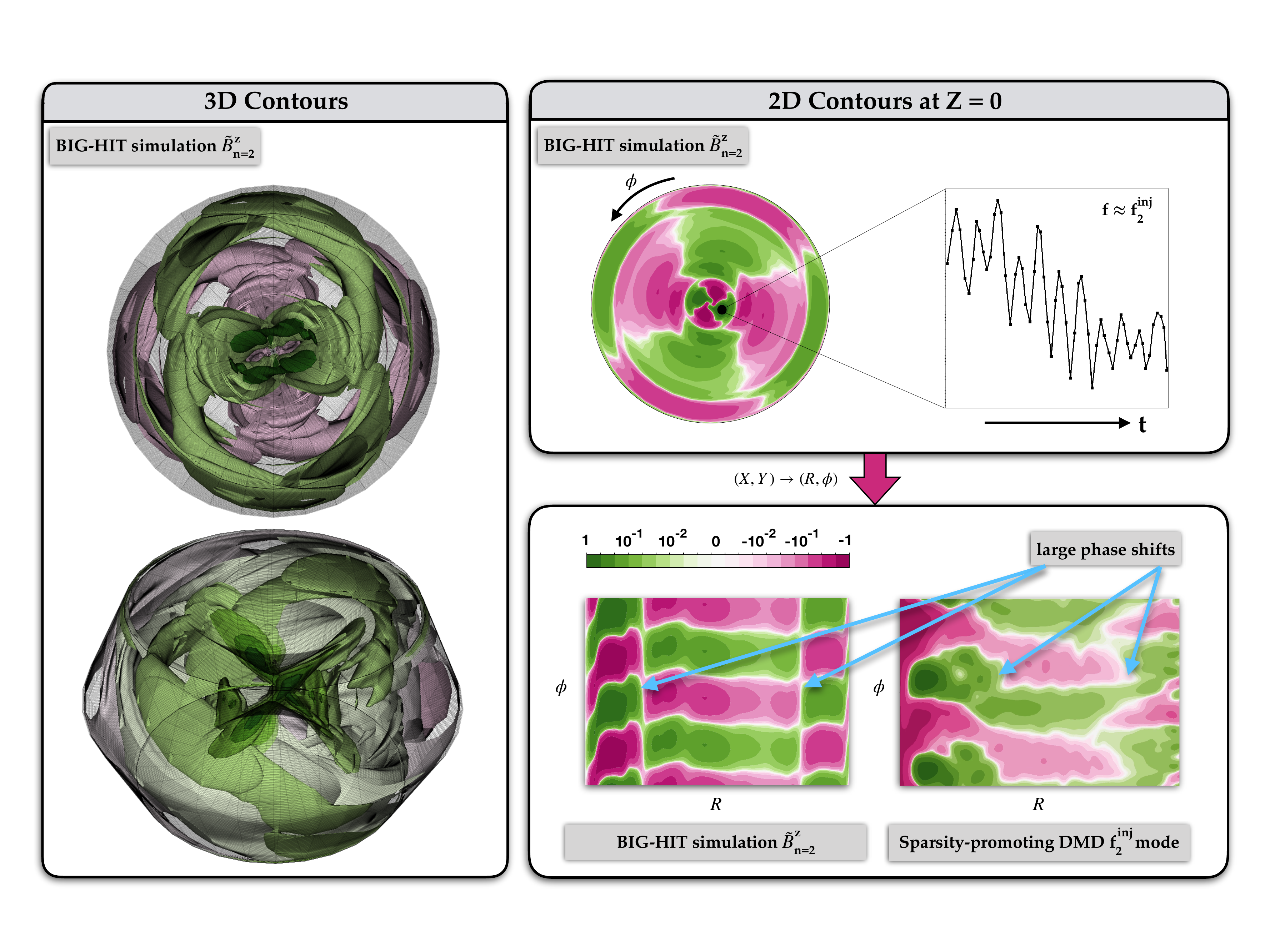}
\end{overpic}
\vspace{-.15in}
    \caption{3D snapshots of $\tilde{B}^z_{n=2}$ from BIG-HIT simulations illustrate a dynamic spiral structure that penetrates down the core and edge of the device, and connects through the injector mouths. Corresponding 2D contour plots at $Z=0$ of the same simulation data indicate that much of this structure is oscillating roughly at $f_2^\text{inj}$, thereby connecting the $f_2^\text{inj}$ mode found in the DMD analysis with a physical structure with the correct $n=2$ dependence. Lastly, a comparison at $Z=0$ between the simulation data $\tilde{B}^z_{n=2}$ and the sparsity-promoting DMD reconstruction using $f_2^\text{inj}$ shows that much of the fine-scale structure can be captured by sparsity-promoting DMD.}
    \label{fig:f2_analysis}
\end{figure*}

To investigate whether or not this mode corresponds to a physical structure, we directly analyze the BIG-HIT simulation $B^z_{n=2}$. We discover a rich, previously unobserved three-dimensional structure in the simulation, shown in Fig.~\ref{fig:f2_analysis}. This structure wraps around the outside of the device by looping through the different injector mouths, and spirals down the core of the device, rotating at $f \approx f_2^\text{inj}$. A comparison  of this structure at $Z=0$ with the DMD reconstruction using only the $f_2^\text{inj}$ mode shows surprising agreement. The reconstruction is able to capture much of the structure observed in the simulation, including the two phase shifts mentioned earlier. 

We have verified that this mode is also present in HIT-SI simulations that evolve the full Hall-MHD equations, as opposed to BIG-HIT, which does not evolve temperature and density. 
These simulations have carefully chosen parameters to match the experiment, and are thus expected to be the most representative of the HIT-SI experiment.  
This is the first identification of a physical and coherent 3D structure in HIT-SI simulation or experiment beyond the dominant injector modes and the spheromak. 

The $f_3^\text{inj}$ mode exhibits mostly odd toroidal mode number dependence and a dominant $n=3$ dependence at $R \approx 0.1$. However, this mode only accounts for approximately $1 \%$ of the total $B_z$ energy and has similar Fourier dependence as the $f_1^\text{inj}$ mode, making it exceedingly difficult to verify in the simulation $B^z_{n=3}$, as was done for the $f_2^\text{inj}$ mode. This difficulty in decomposing the magnetic field into different oscillating and rotating structures is in fact one of the primary motivations for the use of DMD.

A summary of the toroidal and poloidal dependence for the $f_0$, $f_1^\text{inj}$, $f_2^\text{inj}$, and $f_3^\text{inj}$ modes, as well as the $f_\text{kink}$ mode analyzed in the next section, can be found in Fig.~\ref{fig:histograms} for both datasets. The modes exhibit a broad poloidal spectrum that is expected for this device; the surface probes are on the bowtie-shaped boundary, and subsequently tend to observe many poloidal wavenumbers.

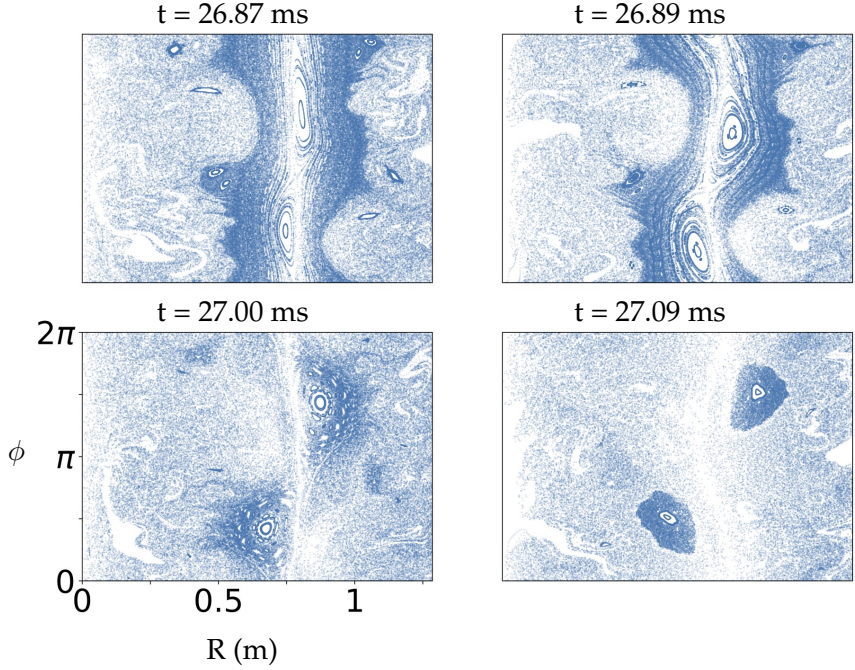
\begin{figure}[!tp]
\vspace{-.1in}
\centering
\begin{overpic}[width=0.8\textwidth]{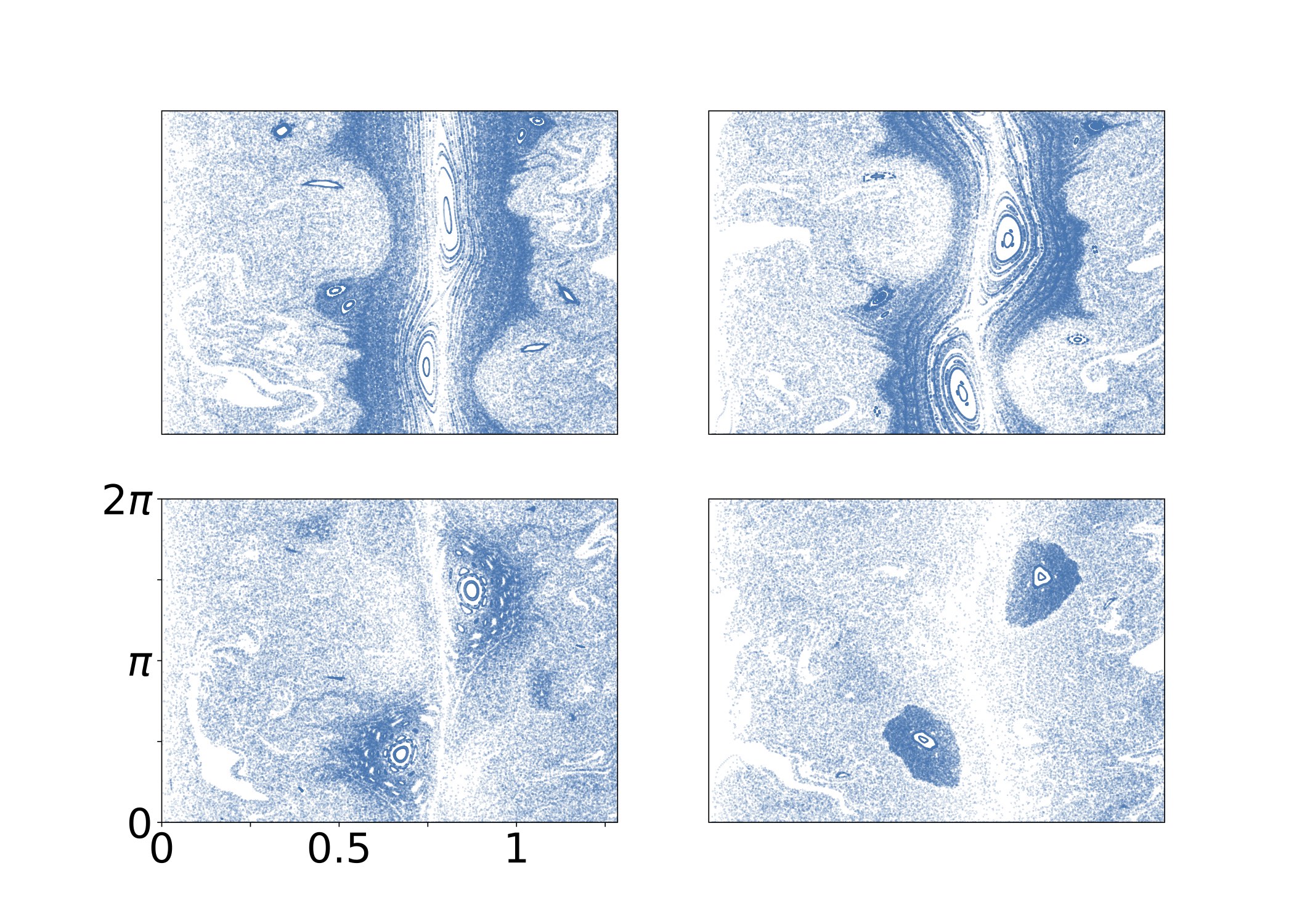}
    \put(25,0){R (m)}
    \put(5,20){$\phi$}
    \put(20,64){t = 26.87 ms}
    \put(62,64){t = 26.89 ms}
    \put(20,34){t = 27.00 ms}
    \put(62,34){t = 27.09 ms}
    \end{overpic}
    \caption{Poincar\'e plots, generated using the full simulation data on the midplane at four consecutive instances, illustrate the evolution of the resistive kink instability.}
    \label{fig:bighit}
\end{figure}

\subsection{Optimized DMD: Kink instability}
Linear MHD stability is of considerable importance in the plasma physics community, especially for confinement devices. While the interpretable models of the previous section allowed for the identification of large-scale physical structures while avoiding overfitting, the optimized DMD is useful for accurate modeling of transient instabilities over smaller time windows.  

In Fig.~\ref{fig:bighit}, Poincar\'e plots from BIG-HIT show closed flux surfaces that exhibit an $n=1$ structure and quasi-periodic sawtooth activity from a $(1,1)$ kink instability, which is consistent with the Kadomtsev or Wesson models~\cite{wesson2011tokamaks}. This simulation is not in the fusion-relevant experimental regime (Lundquist number $\approx 10^8$, plasma temperature $>$ 1 keV) where modifications to the sawtooth behavior from recent work~\cite{krebs2017magnetic} is expected. These modifications predict that sawtooth crashes will not cause large changes to the safety factor on-axis. The BIG-HIT simulation analyzed here used a Lundquist number of approximately $10^6$ and plasma temperature $ = 71$ eV, and showed the existence of large jumps of the on-axis safety factor correlated with the sawtooth activity, providing robust evidence that this is an $(n,m) = (1,1)$ mode well-described by traditional theory. 
The linear growth rate of the resistive $(n,m) = (1,1)$  kink is~\cite{wesson2011tokamaks}
\begin{equation}
    \nu_{11} = \frac{1}{2\pi}\left(\frac{\eta q'(R_1)B_\theta(R_1)}{\mu_0^2 R_1^2\rho}\right)^{\frac{1}{3}}.
\end{equation}
In the equation above, $\rho$ is the mass density, ${R_1 \approx 0.9}\text{ m}$ is the radius which satisfies $q(R_1) \approx 1$, and rough estimates from the previous BIG-HIT analysis yield ${q'(R_1) \approx \Delta q/\Delta R\approx 0.05/0.05 = 1}$ and $B_\theta(R_1) \approx 100$ G. Evaluating with these values gives $\nu_{11} \approx 1000$ $s^{-1}$. 

Optimized DMD captures an $(n,m)=(1,1)$ instability and obtains its growth rate for both datasets in the window $26.8$ ms $\leq t \leq 27.1$ ms.  
This phenomenon is robust for a range of SVD truncation ranks from $10 \leq r \leq 50$; when $r < 10$, the fit does not capture the exponential growth, and when $r > 50$ the poor initial guess results in convergence to a sub-optimal minimum. 
For values of $10 \leq r \leq 50$, $\nu_\text{kink} \approx 600-1100$ $s^{-1}$ using 5120 internal measurements, and $\nu_\text{kink} \approx 600-2000$ $s^{-1}$ using 24 measurements, in excellent agreement with the estimate of $\nu_{11} \approx 1000$ $s^{-1}$.

\begin{figure*}[!tp]
\centering
\begin{overpic}[width=0.95\linewidth]{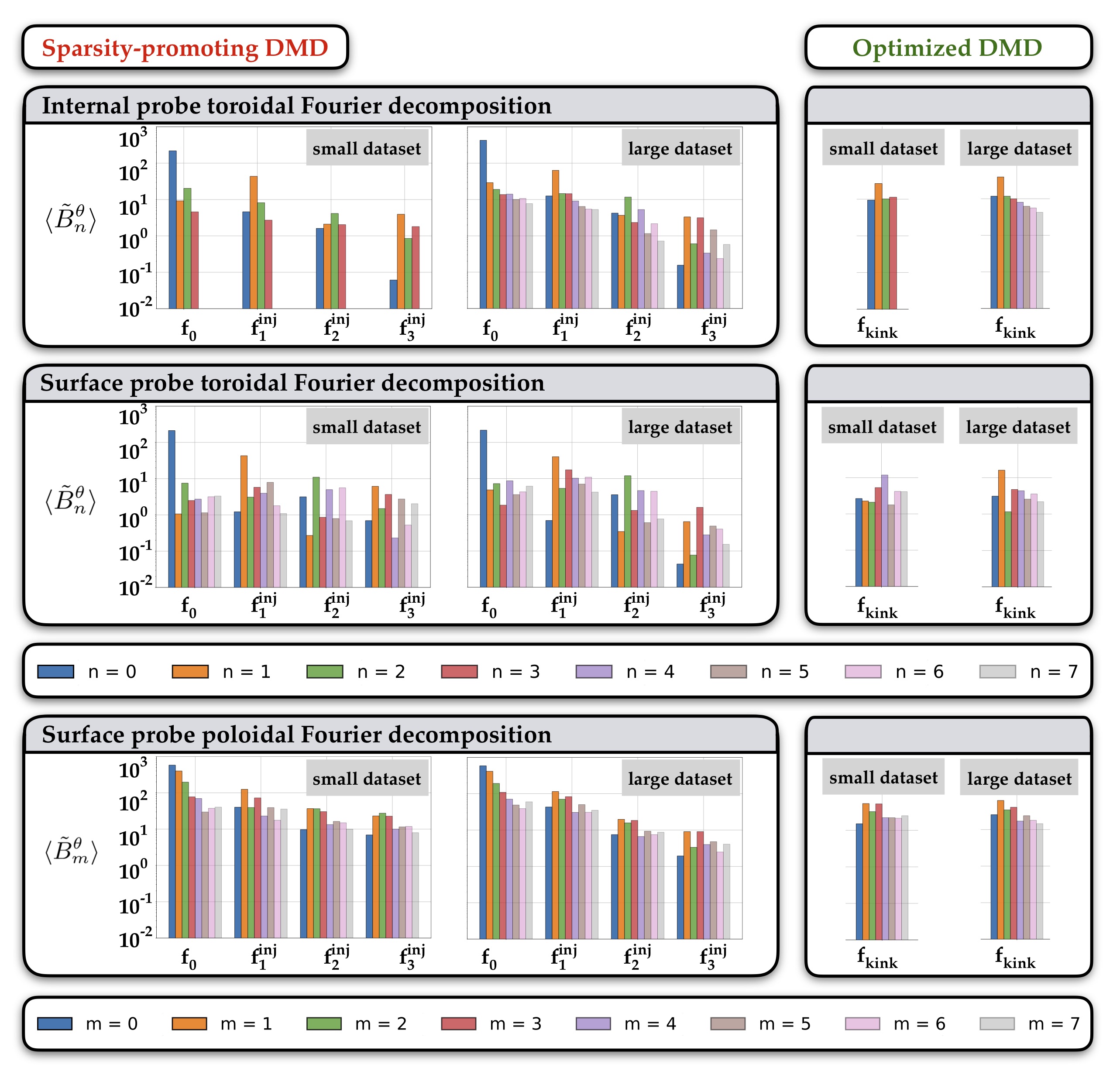}
\end{overpic}
\vspace{-.15in}
    \caption{Averaged toroidal mode content $\langle \tilde{B}^\theta_n \rangle$ of the surface and internal probes for the $f_0$, $f_1^\text{inj}$, $f_2^\text{inj}$, $f_3^\text{inj}$, and $f_\text{kink}$ DMD modes is shown for the two datasets. The surface probe decomposition in the toroidal direction is in excellent agreement with the internal probe data except for the low-resolution kink mode. This illustrates the global structure of $f_0$, $f_1^\text{inj}$, $f_2^\text{inj}$, $f_3^\text{inj}$ and the local spatial structure of the instability near the closed flux region. 
    The $f_0$ mode is almost purely characterized by $n = 0$, reaffirming the physical interpretation of an axisymmetric spheromak. Modes $f_1^\text{inj}$ and $f_3^\text{inj}$ indicate dominant odd $n$ structure, which is expected because the injectors are intentionally driven to produce a mostly $n = 1$ magnetic structure. Interestingly, mode $f_2^\text{inj}$ shows significant mode content in the even $n$ numbers. The instability $f_\text{kink}$ is observed at both low and high resolution to have a $(1,1)$ structure. Magnitude disagreements between the datasets are a result of differences in the number and locations of the probes.
    }
    \label{fig:histograms}
\end{figure*}
To account for some models resulting in many growing modes, only modes with $\nu_j > 100$ $s^{-1}$ are retained; the major spheromak or injector modes are often below this threshold. Any modes oscillating within 1 kHz of the injector frequency are also rejected, in an attempt to control for modes directly driven by the injectors, which are known to have a $n=1$ structure.
The growth rate reported is the weighted average of the remaining growing modes. To validate this approach, the other two kink events observed in the full window {$22.7$ ms $\leq t \leq 28.5$ ms} are analyzed. Again, the results indicate growth rates $\nu_\text{kink} \approx 500-2000$ $s^{-1}$ and similar spatial dependence. 

The toroidal and poloidal Fourier decompositions of the modes, reported in Fig.~\ref{fig:histograms}, indicate an $(n,m)=(1,1)$ structure, and the contour plots in Fig.~\ref{fig:contours} illustrate dominant $n=1$ dependence in the closed flux region. The surface probes indicate an $m=1$ structure of the instability, despite the broad spectrum. With the low-resolution dataset, the surface probes exhibit dominant $n=4$ dependence. Many of the surface probe signals have barely perceptible changes during the transient instability, and thus is reasonable that the spatial dependence of the instability cannot be consistently identified with these probes. The surface probe decomposition exhibits a dominant $n=1$ structure for the high-resolution dataset, which may be consistent with additional probes resulting in a better representation of the magnetic field dependence of the instability. 

Originally, it appeared possible that the injectors directly drive the $n=1$ kinking, but this analysis suggests that plasma generated activity is responsible. Another simulation using a set of four injectors (two each on top and bottom) driving a primarily $n=2$ magnetic structure, indicates periods of closed flux followed by an opening of the flux surfaces by a $n=1$ kink. This lends further evidence to identification of this kink mode independent of the primary injector magnetic configuration. 

\section{Conclusion}\label{Sec:conclusion}
\normalsize
The sparsity-promoting and optimized variants of the dynamic mode decomposition have been shown to enable the discovery of novel magnetic structures from a sparse set of measurements of a driven spheromak.  
 Spatio-temporal modes corresponding to the injector harmonics are identified, along with the characterization of a resistive $(1,1)$ kink instability. 
 Further, the evolution of these modes is accurately captured by a low-rank, interpretable, and linear model, demonstrating the potential for forecasting and real-time control. 
 Importantly, we demonstrate the effectiveness of DMD on data from both the HIT-SI experiment and accompanying BIG-HIT simulations. 

The sparsity-promoting DMD is shown to provide an interpretable and physical model of the major magnetic modes, while avoiding overfitting. This model leads to new discovery of physical coherent structures in the BIG-HIT simulation. The $f_1^\text{inj}$ structure corresponds to the dominant part of the driven injector fields. The $f_2^\text{inj}$ mode on the midplane was used to uncover a previously unobserved 3D structure in the simulation that oscillates at $f_2^\text{inj}$, has $n=2$ toroidal Fourier dependence, and spirals through the injectors near the boundary of the device. 

The optimized DMD demonstrates more accurate signal reconstruction that may be useful for forecasting and characterizing smaller-scale coherent structures. This method fully characterizes a resistive kink instability on a small time window, indicating the ability for robust data-driven  identification of MHD instabilities, with implications for real-time control. These methods, with online training and communication with other machine learning techniques with offline training~\cite{Rea2018,ReaGranetz2018,Montes2019}, could prove useful for disruption mitigation. If both the discovery of interpretable dynamics and the accurate  characterization of instabilities is desired, the authors recommend a joint use of both the sparsity-promoting and optimized DMD algorithms, as illustrated here. 

While this paper has focused on magnetic measurements from a number of simple probes, in principle, these methods only rely on a set of sparse experimental measurements of any relevant plasma quantity.  
Thus, they should be highly applicable to a wealth of different diagnostics and plasmas spanning much of the possible parameter space. Although all reduced-order methods discussed here result in global spatial modes, the analysis can be restricted to small-scale spatial structures by using a small number of nearby probes. Despite the localization of the resistive kink instability near the closed flux, it was successfully identified with these methods by radially averaging over the internal probe arrays, and visually confirming the $n=1$ structure in the closed flux region. To capture transient modes such as the resistive kink identified in this paper, the methods can be applied on a small time window. This flexibility and generality make DMD an excellent choice for the discovery of coherent plasma structures and instabilities, and subsequent attempts to control them.

There are a number of important avenues of future research that are suggested by this analysis. Zonal flows are ubiquitous, and their spatio-temporal structures are important for turbulent transport. A detailed analysis of the low frequency global potential oscillations on the TJ-II stellarator indicated the spatial structure had high sensitivity to the mean electric field profile~\cite{Kobayashi2019}. A data-driven DMD model for the evolution of the spatial structure of the zonal flows may facilitate future control over the turbulent transport properties mediated by this mode. 

For plasmas with important small-scale and transient turbulent structures, the biorthogonal wavelet decomposition and reduced-order methods based on wavelets~\cite{farge2015wavelet}, such as multi-resolution DMD~\cite{kutz2016multiresolution} and multi-resolution biorthogonal decomposition~\cite{mendez2019multi}, show significant promise. These would be ideal for the analysis of coherent structures with small spatial and temporal correlation lengths; for instance, quasi-coherent modes of this sort appear to be universal in the edge of Alcator C-Mod plasmas~\cite{golfinopoulos2019wavelet}. The DMD framework has several other extensions that may improve the analysis of complex experimental plasmas, which often have limited measurements, complex multi-scale dynamics, and actuation. Another promising model for mapping nonlinear dynamics onto approximate linear representations is the Hankel alternative view of Koopman (HAVOK) analysis~\cite{brunton2017chaos}, which utilizes time-delay coordinates to enrich the sensor space. 

Discovering the underlying coherent structures also facilitates control~\cite{Brunton2015amr}. There are many unsolved open and closed loop control problems in experimental plasma devices~\cite{ariola2008magnetic,Levesque2017}. The DMD algorithm has been previously extended to decompose signals while disambiguating the system dynamics from the effects of external forcing. This is ideal for discovering and then controlling plasma-generated dynamics. In fact, DMD with control (DMDc) has shown significant improvement over the traditional method on externally forced systems~\cite{proctor2016dynamic}. For the purposes of this work, it was found that performance was similar when injector current waveforms were treated as actuators. However, for other experimental devices, accounting for external actuation may significantly improve discovery of plasma dynamics.  

\section{Acknowledgements}
The analysis performed in this work was funded by the Air Force Office of Scientific Research (AFOSR FA9550-18-1-0200). The experimental and simulation data comes from work supported by the U.S. Department of Energy, Office of Science, Office of Fusion Energy Sciences, under award numbers DE-FG02-96ER54361 and {DE-SC}0016256. Simulations presented here used resources of the National Energy Research Scientific Computing Center, supported by the Office of Science of the U.S. Department of Energy under contract number DE-AC02-05CH11231. This work also used the PSC Bridges supercomputer under allocation ID TG-PHY180064 through the Extreme Science and Engineering Discovery Environment (XSEDE)~\cite{xsede}, which is supported by National Science Foundation grant number ACI-1548562. Data analysis was facilitated through the use of advanced computational, storage, and networking infrastructure provided by the Hyak supercomputer system and funded by the student technology fund (STF) at the University of Washington.


\bibliographystyle{plain}
 \begin{spacing}{.9}
 \small{
 \setlength{\bibsep}{6.5pt}


\begin{thebibliography}{10}

\bibitem{akcay2013extended}
Cihan Akcay.
\newblock {\em Extended magnetohydrodynamic simulations of the helicity
  injected torus ({HIT-SI}) spheromak experiment with the {NIMROD} code}.
\newblock PhD thesis, {U}niversity of {W}ashington, {S}eattle, 2013.

\bibitem{antolin2017observational}
Patrick Antolin, Ineke De~Moortel, Tom Van~Doorsselaere, and Takaaki Yokoyama.
\newblock Observational signatures of transverse magnetohydrodynamic waves and
  associated dynamic instabilities in coronal flux tubes.
\newblock {\em The Astrophysical Journal}, 836(2):219, 2017.

\bibitem{ariola2008magnetic}
Marco Ariola and Alfredo Pironti.
\newblock {\em Magnetic control of tokamak plasmas}, volume 187.
\newblock Springer, 2008.

\bibitem{askham2018variable}
Travis Askham and J~Nathan Kutz.
\newblock Variable projection methods for an optimized dynamic mode
  decomposition.
\newblock {\em SIAM Journal on Applied Dynamical Systems}, 17(1):380--416,
  2018.

\bibitem{Bagheri2014pof}
Shervin Bagheri.
\newblock Effects of weak noise on oscillating flows: Linking quality factor,
  {F}loquet modes, and {K}oopman spectrum.
\newblock {\em Physics of Fluids}, 26(9):094104, 2014.

\bibitem{belonohy2008systematic}
E~Belonohy, G~Pokol, K~McCormick, G~Papp, S~Zoletnik, and W7-AS Team.
\newblock A systematic study of the quasi-coherent mode in the high density
  {H}-mode regime of {W}endelstein 7-{AS}.
\newblock In {\em AIP Conference Proceedings}, volume 993, pages 39--42. AIP,
  2008.

\bibitem{Benner2015siamreview}
Peter Benner, Serkan Gugercin, and Karen Willcox.
\newblock A survey of projection-based model reduction methods for parametric
  dynamical systems.
\newblock {\em SIAM review}, 57(4):483--531, 2015.

\bibitem{benson2013direct}
Austin~R Benson, David~F Gleich, and James Demmel.
\newblock Direct {QR} factorizations for tall-and-skinny matrices in
  {M}ap{R}educe architectures.
\newblock In {\em 2013 IEEE international conference on big data}, pages
  264--272. IEEE, 2013.

\bibitem{Brunton2015amr}
S.~L. Brunton and B.~R. Noack.
\newblock Closed-loop turbulence control: Progress and challenges.
\newblock {\em Applied Mechanics Reviews}, 67:050801--1--050801--48, 2015.

\bibitem{Brunton2015jcd}
S.~L. Brunton, J.~L. Proctor, J.~H. Tu, and J.~N. Kutz.
\newblock Compressed sensing and dynamic mode decomposition.
\newblock {\em Journal of Computational Dynamics}, 2(2):165--191, 2015.

\bibitem{brunton2017chaos}
Steven~L Brunton, Bingni~W Brunton, Joshua~L Proctor, Eurika Kaiser, and
  J~Nathan Kutz.
\newblock Chaos as an intermittently forced linear system.
\newblock {\em Nature communications}, 8(1):19, 2017.

\bibitem{brunton2019data}
Steven~L Brunton and J~Nathan Kutz.
\newblock {\em Data-driven science and engineering: Machine learning, dynamical
  systems, and control}.
\newblock Cambridge University Press, 2019.

\bibitem{Brunton2020arfm}
Steven~L. Brunton, Bernd~R. Noack, and Petros Koumoutsakos.
\newblock Machine learning for fluid mechanics.
\newblock {\em to appear in Annual Review of Fluid Mechanics (arXiv preprint
  arXiv: 1905.11075)}, 52, 2020.

\bibitem{burrell2002quiescent}
KH~Burrell, Max~E Austin, DP~Brennan, JC~DeBoo, EJ~Doyle, P~Gohil,
  CM~Greenfield, RJ~Groebner, LL~Lao, TC~Luce, et~al.
\newblock Quiescent {H}-mode plasmas in the {DIII-D} tokamak.
\newblock {\em Plasma Physics and Controlled Fusion}, 44(5A):A253, 2002.

\bibitem{byrne2017study}
Patrick~James Byrne.
\newblock {\em Study of External Kink Modes in Shaped HBT-EP Plasmas}.
\newblock PhD thesis, Columbia University, 2017.

\bibitem{Carlberg2017jcp}
Kevin Carlberg, Matthew Barone, and Harbir Antil.
\newblock Galerkin v. least-squares petrov--galerkin projection in nonlinear
  model reduction.
\newblock {\em Journal of Computational Physics}, 330:693--734, 2017.

\bibitem{Dawson2016ef}
Scott~TM Dawson, Maziar~S Hemati, Matthew~O Williams, and Clarence~W Rowley.
\newblock Characterizing and correcting for the effect of sensor noise in the
  dynamic mode decomposition.
\newblock {\em Experiments in Fluids}, 57(3):1--19, 2016.

\bibitem{dudok1994biorthogonal}
T~Dudok~de Wit, A-L Pecquet, J-C Vallet, and R~Lima.
\newblock The biorthogonal decomposition as a tool for investigating
  fluctuations in plasmas.
\newblock {\em Physics of plasmas}, 1(10):3288--3300, 1994.

\bibitem{jetsawtooth}
A.~W. Edwards, D.~J. Campbell, W.~W. Engelhardt, H.~U. Fahrbach, R.~D. Gill,
  R.~S. Granetz, S.~Tsuji, B.~J.~D. Tubbing, A.~Weller, J.~Wesson, and
  D.~Zasche.
\newblock Rapid collapse of a plasma sawtooth oscillation in the {JET} tokamak.
\newblock {\em Phys. Rev. Lett.}, 57:210--213, {J}uly 1986.

\bibitem{farge2015wavelet}
Marie Farge and Kai Schneider.
\newblock Wavelet transforms and their applications to {MHD} and plasma
  turbulence: a review.
\newblock {\em Journal of Plasma Physics}, 81(6), 2015.

\bibitem{foullon2011magnetic}
Claire Foullon, Erwin Verwichte, Valery~M Nakariakov, Katariina Nykyri, and
  Charles~J Farrugia.
\newblock Magnetic {K}elvin-{H}elmholtz instability at the sun.
\newblock {\em The Astrophysical Journal Letters}, 729(1):L8, 2011.

\bibitem{galperti2014development}
C~Galperti, C~Marchetto, E~Alessi, D~Minelli, M~Mosconi, F~Belli, L~Boncagni,
  A~Botrugno, P~Buratti, B~Esposito, et~al.
\newblock Development of real-time {MHD} markers based on biorthogonal
  decomposition of signals from {M}irnov coils.
\newblock {\em Plasma Physics and Controlled Fusion}, 56(11):114012, 2014.

\bibitem{ghadimi2014optimal}
Euhanna Ghadimi, Andr{\'e} Teixeira, Iman Shames, and Mikael Johansson.
\newblock Optimal parameter selection for the alternating direction method of
  multipliers ({ADMM}): quadratic problems.
\newblock {\em IEEE Transactions on Automatic Control}, 60(3):644--658, 2014.

\bibitem{goertz1979magnetosphere}
CK~Goertz and RW~Boswell.
\newblock Magnetosphere-ionosphere coupling.
\newblock {\em Journal of Geophysical Research: Space Physics},
  84(A12):7239--7246, 1979.

\bibitem{golfinopoulos2014external}
T~Golfinopoulos, B~LaBombard, RR~Parker, W~Burke, E~Davis, R~Granetz,
  M~Greenwald, J~Irby, R~Leccacorvi, E~Marmar, et~al.
\newblock External excitation of a short-wavelength fluctuation in the
  {A}lcator {C-M}od edge plasma and its relationship to the quasi-coherent
  mode.
\newblock {\em Physics of Plasmas}, 21(5):056111, 2014.

\bibitem{golfinopoulos2019wavelet}
Theodore Golfinopoulos.
\newblock The wavelet nature of persistent edge fluctuations observed on
  {A}lcator {C-M}od.
\newblock {\em Bulletin of the American Physical Society}, 2019.

\bibitem{DIII-D}
C.~M. Greenfield, K.~H. Burrell, J.~C. DeBoo, E.~J. Doyle, B.~W. Stallard,
  E.~J. Synakowski, C.~Fenzi, P.~Gohil, R.~J. Groebner, L.~L. Lao, M.~A.
  Makowski, G.~R. McKee, R.~A. Moyer, C.~L. Rettig, T.~L. Rhodes, R.~I.
  Pinsker, G.~M. Staebler, W.~P. West, and the DIII-D~Team.
\newblock Quiescent double barrier regime in the {DIII-D} tokamak.
\newblock {\em Phys. Rev. Lett.}, 86:4544--4547, May 2001.

\bibitem{Gueniat2015pof}
F.~Gueniat, L.~Mathelin, and L.~Pastur.
\newblock A dynamic mode decomposition approach for large and arbitrarily
  sampled systems.
\newblock {\em Physics of Fluids}, 27(2):025113, 2015.

\bibitem{hansen2015simulation}
C~Hansen, G~Marklin, B~Victor, C~Akcay, and T~Jarboe.
\newblock Simulation of injector dynamics during steady inductive helicity
  injection current drive in the {HIT-SI} experiment.
\newblock {\em Physics of Plasmas}, 22(4):042505, 2015.

\bibitem{hansen2015numerical}
C~Hansen, B~Victor, K~Morgan, T~Jarboe, A~Hossack, G~Marklin, BA~Nelson, and
  D~Sutherland.
\newblock Numerical studies and metric development for validation of
  magnetohydrodynamic models on the {HIT-SI} experiment.
\newblock {\em Physics of Plasmas}, 22(5):056105, 2015.

\bibitem{Hemati2014pof}
Maziar~S Hemati, Matthew~O Williams, and Clarence~W Rowley.
\newblock Dynamic mode decomposition for large and streaming datasets.
\newblock {\em Physics of Fluids (1994-present)}, 26(11):111701, 2014.

\bibitem{hossack2015study}
Aaron Hossack.
\newblock {\em A study of plasma dynamics in {HIT-SI} using ion Doppler
  spectroscopy}.
\newblock PhD thesis, {U}niversity of {W}ashington, {S}eattle, 2015.

\bibitem{hossack2017plasma}
AC~Hossack, TR~Jarboe, RN~Chandra, KD~Morgan, DA~Sutherland, JM~Penna,
  CJ~Everson, and BA~Nelson.
\newblock Plasma response to sustainment with imposed-dynamo current drive in
  {HIT-SI} and {HIT-SI3}.
\newblock {\em Nuclear Fusion}, 57(7):076026, 2017.

\bibitem{jarboe1994review}
Thomas~R Jarboe.
\newblock Review of spheromak research.
\newblock {\em Plasma Physics and Controlled Fusion}, 36(6):945, 1994.

\bibitem{jarboe2006spheromak}
TR~Jarboe, WT~Hamp, GJ~Marklin, BA~Nelson, RG~O’Neill, AJ~Redd, PE~Sieck,
  RJ~Smith, and JS~Wrobel.
\newblock Spheromak formation by steady inductive helicity injection.
\newblock {\em Physical review letters}, 97(11):115003, 2006.

\bibitem{jarboe2012imposed}
TR~Jarboe, BS~Victor, BA~Nelson, CJ~Hansen, C~Akcay, DA~Ennis, NK~Hicks,
  AC~Hossack, GJ~Marklin, and RJ~Smith.
\newblock Imposed-dynamo current drive.
\newblock {\em Nuclear Fusion}, 52(8):083017, 2012.

\bibitem{jovanovic2014sparsity}
Mihailo~R Jovanovi{\'c}, Peter~J Schmid, and Joseph~W Nichols.
\newblock Sparsity-promoting dynamic mode decomposition.
\newblock {\em Physics of Fluids}, 26(2):024103, 2014.

\bibitem{kamiya2004high}
K~Kamiya, M~Bakhtiari, S~Kasai, H~Kawashima, Y~Kusama, Y~Miura, H~Ogawa,
  N~Oyama, M~Sato, K~Shinohara, et~al.
\newblock High recycling steady {H}-mode regime in the {JFT-2M} tokamak.
\newblock {\em Plasma physics and controlled fusion}, 46(5A):A157, 2004.

\bibitem{khalil2002nonlinear}
Hassan~K Khalil.
\newblock Nonlinear systems.
\newblock {\em Upper Saddle River}, 2002.

\bibitem{klus2017data}
Stefan Klus, Feliks N{\"u}ske, P{\'e}ter Koltai, Hao Wu, Ioannis Kevrekidis,
  Christof Sch{\"u}tte, and Frank No{\'e}.
\newblock Data-driven model reduction and transfer operator approximation.
\newblock {\em Journal of Nonlinear Science}, 2018.

\bibitem{Kobayashi2019}
T.~Kobayashi, U.~Losada, B.~Liu, T.~Estrada, B.Ph. van Milligen,
  R.~Gerr{\'{u}}, M.~Sasaki, and C.~Hidalgo.
\newblock Frequency and plasma condition dependent spatial structure of low
  frequency global potential oscillations in the {TJ}-{II} stellarator.
\newblock {\em Nuclear Fusion}, 59(4):044006, mar 2019.

\bibitem{krebs2017magnetic}
I~Krebs, SC~Jardin, S~G{\"u}nter, K~Lackner, M~Hoelzl, E~Strumberger, and
  N~Ferraro.
\newblock Magnetic flux pumping in 3{D} nonlinear magnetohydrodynamic
  simulations.
\newblock {\em Physics of Plasmas}, 24(10):102511, 2017.

\bibitem{Kutz2016book}
J.~N. Kutz, S.~L. Brunton, B.~W. Brunton, and J.~L. Proctor.
\newblock {\em Dynamic Mode Decomposition: Data-Driven Modeling of Complex
  Systems}.
\newblock SIAM, 2016.

\bibitem{kutz2016multiresolution}
J~Nathan Kutz, Xing Fu, and Steven~L Brunton.
\newblock Multiresolution dynamic mode decomposition.
\newblock {\em SIAM Journal on Applied Dynamical Systems}, 15(2):713--735,
  2016.

\bibitem{levenberg1944method}
Kenneth Levenberg.
\newblock A method for the solution of certain non-linear problems in least
  squares.
\newblock {\em Quarterly of applied mathematics}, 2(2):164--168, 1944.

\bibitem{Levesque2017}
J.P. Levesque, J.W. Brooks, M.C. Abler, J.~Bialek, P.J. Byrne, C.J. Hansen,
  P.E. Hughes, M.E. Mauel, G.A. Navratil, and D.J. Rhodes.
\newblock Measurement of scrape-off-layer current dynamics during {MHD}
  activity and disruptions in {HBT}-{EP}.
\newblock {\em Nuclear Fusion}, 57(8):086035, {J}uly 2017.

\bibitem{marquardt1963algorithm}
Donald~W Marquardt.
\newblock An algorithm for least-squares estimation of nonlinear parameters.
\newblock {\em Journal of the society for Industrial and Applied Mathematics},
  11(2):431--441, 1963.

\bibitem{mendez2019multi}
MA~Mendez, M~Balabane, and J-M Buchlin.
\newblock Multi-scale proper orthogonal decomposition of complex fluid flows.
\newblock {\em Journal of Fluid Mechanics}, 870:988--1036, 2019.

\bibitem{Mezic2005nd}
Igor Mezi{\'c}.
\newblock Spectral properties of dynamical systems, model reduction and
  decompositions.
\newblock {\em Nonlinear Dynamics}, 41(1-3):309--325, 2005.

\bibitem{Mezic2013arfm}
Igor Mezic.
\newblock Analysis of fluid flows via spectral properties of the {K}oopman
  operator.
\newblock {\em Annual Review of Fluid Mechanics}, 45:357--378, 2013.

\bibitem{Montes2019}
K.~J. Montes, C.~Rea, R.~S. Granetz, R.~A. Tinguely, N.~Eidietis, O.~M.
  Meneghini, D.~L. Chen, B.~Shen, B.~J. Xiao, K.~Erickson, and M.~D. Boyer.
\newblock Machine learning for disruption warnings on {A}lcator {C}-{M}od,
  {DIII}-{D}, and {EAST}.
\newblock {\em Nuclear Fusion}, 59(9):096015, July 2019.

\bibitem{morgan2018finite}
Kyle Morgan.
\newblock {\em Finite-beta simulations of {HIT-SI} and {HIT-SI}3 using the
  {NIMROD} code}.
\newblock PhD thesis, {U}niversity of {W}ashington, {S}eattle, 2018.

\bibitem{morgan2019formation}
Kyle Morgan, Thomas Jarboe, and Cihan Akcay.
\newblock Formation of closed flux surfaces in spheromaks sustained by steady
  inductive helicity injection.
\newblock {\em Nuclear Fusion}, 59(6):066037, 2019.

\bibitem{Noack2003jfm}
B.~R. Noack, K.~Afanasiev, M.~Morzynski, G.~Tadmor, and F.~Thiele.
\newblock A hierarchy of low-dimensional models for the transient and
  post-transient cylinder wake.
\newblock {\em Journal of Fluid Mechanics}, 497:335--363, 2003.

\bibitem{Noack2016jfm}
Bernd~R Noack, Witold Stankiewicz, Marek Morzynski, and Peter~J Schmid.
\newblock Recursive dynamic mode decomposition of a transient cylinder wake.
\newblock {\em Journal of Fluid Mechanics}, 809:843--872, 2016.

\bibitem{noe2013variational}
Frank No{\'e} and Feliks Nuske.
\newblock A variational approach to modeling slow processes in stochastic
  dynamical systems.
\newblock {\em Multiscale Modeling \ Simulation}, 11(2):635--655, 2013.

\bibitem{nuske2014jctc}
Feliks N{\"u}ske, Bettina~G Keller, Guillermo P{\'e}rez-Hern{\'a}ndez,
  Antonia~SJS Mey, and Frank No{\'e}.
\newblock Variational approach to molecular kinetics.
\newblock {\em Journal of chemical theory and computation}, 10(4):1739--1752,
  2014.

\bibitem{pandya}
Mihir Pandya.
\newblock {\em Low edge safety factor disruptions in the Compact Toroidal
  Hybrid: Operation in the low-q regime, passive disruption avoidance and the
  nature of {MHD} precursors}.
\newblock PhD thesis, {A}uburn {U}niversity, 2016.

\bibitem{proctor2016dynamic}
Joshua~L Proctor, Steven~L Brunton, and J~Nathan Kutz.
\newblock Dynamic mode decomposition with control.
\newblock {\em SIAM Journal on Applied Dynamical Systems}, 15(1):142--161,
  2016.

\bibitem{ReaGranetz2018}
C.~Rea and R.~S. Granetz.
\newblock Exploratory machine learning studies for disruption prediction using
  large databases on {DIII-D}.
\newblock {\em Fusion Science and Technology}, 74(1-2):89--100, 2018.

\bibitem{Rea2018}
C.~Rea, R.~S. Granetz, K.~Montes, R.~A. Tinguely, N.~Eidietis, J.~M. Hanson,
  and B.~Sammuli.
\newblock Disruption prediction investigations using machine learning tools on
  {DIII}-{D} and {A}lcator {C}-{M}od.
\newblock {\em Plasma Physics and Controlled Fusion}, 60(8):084004, {J}une
  2018.

\bibitem{Rowley2009jfm}
C.~W. Rowley, I.\ Mezi\'c, S.\ Bagheri, P.\ Schlatter, and D.S. Henningson.
\newblock Spectral analysis of nonlinear flows.
\newblock {\em J.\ Fluid Mech.}, 645:115--127, 2009.

\bibitem{Rowley2017arfm}
Clarence~W Rowley and Scott~TM Dawson.
\newblock Model reduction for flow analysis and control.
\newblock {\em Annual Review of Fluid Mechanics}, 49:387--417, 2017.

\bibitem{Sasaki2019}
M~Sasaki, Y~Kawachi, R~O Dendy, H~Arakawa, N~Kasuya, F~Kin, K~Yamasaki, and
  S~Inagaki.
\newblock Using dynamical mode decomposition to extract the limit cycle
  dynamics of modulated turbulence in a plasma simulation.
\newblock {\em Plasma Physics and Controlled Fusion}, 61(11):112001, {O}ct.
  2019.

\bibitem{schmid2010dynamic}
Peter~J Schmid.
\newblock Dynamic mode decomposition of numerical and experimental data.
\newblock {\em Journal of fluid mechanics}, 656:5--28, 2010.

\bibitem{sovinec2004nonlinear}
CR~Sovinec, AH~Glasser, TA~Gianakon, DC~Barnes, RA~Nebel, SE~Kruger,
  DD~Schnack, SJ~Plimpton, A~Tarditi, MS~Chu, et~al.
\newblock Nonlinear magnetohydrodynamics simulation using high-order finite
  elements.
\newblock {\em Journal of Computational Physics}, 195(1):355--386, 2004.

\bibitem{cassini2}
A.~H. Sulaiman, W.~S. Kurth, G.~B. Hospodarsky, T.~F. Averkamp, S.-Y. Ye, J.~D.
  Menietti, W.~M. Farrell, D.~A. Gurnett, A.~M. Persoon, M.~K. Dougherty, and
  G.~J. Hunt.
\newblock Enceladus auroral hiss emissions during {C}assini's grand finale.
\newblock {\em Geophysical Research Letters}, 45(15):7347--7353, 2018.

\bibitem{cassini1}
A.~H. Sulaiman, W.~S. Kurth, A.~M. Persoon, J.~D. Menietti, W.~M. Farrell,
  S.-Y. Ye, G.~B. Hospodarsky, D.~A. Gurnett, and L.~Z. Hadid.
\newblock Intense {H}armonic {E}missions observed in {S}aturn's ionosphere.
\newblock {\em Geophysical Research Letters}, 44(24):12,049--12,056, 2017.

\bibitem{swanson2003plasma}
Donald~Gary Swanson.
\newblock {\em Plasma waves}.
\newblock CRC Press, 2003.

\bibitem{Taira2017aiaa}
Kunihiko Taira, Steven~L Brunton, Scott Dawson, Clarence~W Rowley, Tim
  Colonius, Beverley~J McKeon, Oliver~T Schmidt, Stanislav Gordeyev, Vassilios
  Theofilis, and Lawrence~S Ukeiley.
\newblock Modal analysis of fluid flows: An overview.
\newblock {\em AIAA Journal}, 55(12):4013--4041, 2017.

\bibitem{Takeishi2017nips}
Naoya Takeishi, Yoshinobu Kawahara, and Takehisa Yairi.
\newblock Learning {K}oopman invariant subspaces for dynamic mode
  decomposition.
\newblock In {\em Advances in Neural Information Processing Systems}, pages
  1130--1140, 2017.

\bibitem{taylor2018dynamic}
Roy Taylor, J~Nathan Kutz, Kyle Morgan, and Brian~A Nelson.
\newblock Dynamic mode decomposition for plasma diagnostics and validation.
\newblock {\em Review of Scientific Instruments}, 89(5):053501, 2018.

\bibitem{teisberg2000valuation}
Thomas~J Teisberg and Rodney~F Weiher.
\newblock Valuation of geomagnetic storm forecasts: an estimate of the net
  economic benefits of a satellite warning system.
\newblock {\em Journal of Policy Analysis and Management}, 19(2):329--334,
  2000.

\bibitem{Towne2018jfm}
Aaron Towne, Oliver~T Schmidt, and Tim Colonius.
\newblock Spectral proper orthogonal decomposition and its relationship to
  dynamic mode decomposition and resolvent analysis.
\newblock {\em Journal of Fluid Mechanics}, 847:821--867, 2018.

\bibitem{xsede}
J.~Towns, T.~Cockerill, M.~Dahan, I.~Foster, K.~Gaither, A.~Grimshaw,
  V.~Hazlewood, S.~Lathrop, D.~Lifka, G.~D. Peterson, R.~Roskies, J.~R. Scott,
  and N.~Wilkins-Diehr.
\newblock {XSEDE}: Accelerating scientific discovery.
\newblock {\em Computing in Science \& Engineering}, 16(5):62--74,
  {S}ept.-{O}ct. 2014.

\bibitem{Tu2014jcd}
J.~H. Tu, C.~W. Rowley, D.~M. Luchtenburg, S.~L. Brunton, and J.~N. Kutz.
\newblock On dynamic mode decomposition: theory and applications.
\newblock {\em Journal of Computational Dynamics}, 1(2):391--421, 2014.

\bibitem{Tu2014ef}
Jonathan~H Tu, Clarence~W Rowley, J~Nathan Kutz, and Jessica~K Shang.
\newblock Spectral analysis of fluid flows using sub-{N}yquist-rate {PIV} data.
\newblock {\em Experiments in Fluids}, 55(9):1--13, 2014.

\bibitem{victor2015development}
BS~Victor, C~Akcay, CJ~Hansen, TR~Jarboe, BA~Nelson, and KD~Morgan.
\newblock Development of validation metrics using biorthogonal decomposition
  for the comparison of magnetic field measurements.
\newblock {\em Plasma Physics and Controlled Fusion}, 57(4):045010, 2015.

\bibitem{victor2014sustained}
BS~Victor, TR~Jarboe, CJ~Hansen, C~Akcay, KD~Morgan, AC~Hossack, and BA~Nelson.
\newblock Sustained spheromaks with ideal n= 1 kink stability and pressure
  confinement.
\newblock {\em Physics of Plasmas}, 21(8):082504, 2014.

\bibitem{weisen1989mode}
H~Weisen, K~Appert, GG~Borg, B~Joye, AJ~Knight, JB~Lister, and J~Vaclavik.
\newblock Mode conversion to the kinetic alfv{\'e}n wave in low-frequency
  heating experiments in the {TCA} tokamak.
\newblock {\em Physical review letters}, 63(22):2476, 1989.

\bibitem{wesson2011tokamaks}
John Wesson and David~J Campbell.
\newblock {\em Tokamaks}, volume 149.
\newblock Oxford university press, 2011.

\bibitem{williams2015data}
Matthew~O Williams, Ioannis~G Kevrekidis, and Clarence~W Rowley.
\newblock A data--driven approximation of the {k}oopman operator: Extending
  dynamic mode decomposition.
\newblock {\em Journal of Nonlinear Science}, 25(6):1307--1346, 2015.

\bibitem{wrobel2011study}
Jonathan~Scott Wrobel.
\newblock {\em A study of {HIT-SI} plasma dynamics using surface magnetic field
  measurements}.
\newblock University of Washington, 2011.

\end{thebibliography}
 }
 \end{spacing}

\end{document}